\documentclass[12pt,a4paper]{article}

\usepackage[nosort]{cite}
\usepackage[hyperref,bulletsep]{collect}
\usepackage{graphicx}
\usepackage{pst-all}
\usepackage{bbm}
\usepackage{amsmath}
\usepackage{amssymb}
\usepackage{subfigure}

\makeatletter \@addtoreset{equation}{section} \makeatother

\makeatletter
\let\old@startsection=\@startsection
\let\oldl@section=\l@section
\renewcommand{\@startsection}[6]{\old@startsection{#1}{#2}{#3}{#4}{#5}{#6\mathversion{bold}}}
\renewcommand{\l@section}[2]{\oldl@section{\mathversion{bold}#1}{#2}}
\makeatother

\makeatletter
\let\old@makecaption=\@makecaption
\def\@makecaption{\small\old@makecaption}
\makeatother

\renewcommand{\thefootnote}{\arabic{footnote}}
\let\oldPhi=\Phi
\let\oldPsi=\Psi
\let\oldGamma=\Gamma
\let\oldDelta=\Delta
\let\oldSigma=\Sigma
\let\oldTheta=\Theta
\let\oldPi=\Pi
\let\oldUpsilon=\Upsilon
\renewcommand{\Phi}{\mathnormal{\oldPhi}}
\renewcommand{\Psi}{\mathnormal{\oldPsi}}
\renewcommand{\Gamma}{\mathnormal{\oldGamma}}
\renewcommand{\Sigma}{\mathnormal{\oldSigma}}
\renewcommand{\Delta}{\mathnormal{\oldDelta}}
\renewcommand{\Theta}{\mathnormal{\oldTheta}}
\renewcommand{\Pi}{\mathnormal{\oldPi}}
\renewcommand{\Upsilon}{\mathnormal{\oldUpsilon}}

\newcommand{\Amp}{\mathcal{A}}

\newcommand{\csch}{\operatorname{csch}}
\newcommand{\sech}{\operatorname{sech}}
\newcommand{\arcsinh}{\operatorname{arcsinh}}

\newcommand{\order}{\mathcal{O}}

\newcommand{\Sphere}{S}  
\newcommand{\AdS}{\mathrm{AdS}}

\ifx\genfrac\sdflkaj

\else

\fi
\newcommand{\sfrac}[2]{{\textstyle\frac{#1}{#2}}}
\newcommand{\half}{\sfrac{1}{2}}

\newcommand{\Half}{\frac{1}{2}}

\newcommand{\matr}[2]{\left(\begin{array}{#1}#2\end{array}\right)}
\newcommand{\alg}[1]{\mathfrak{#1}}
\newcommand{\grp}[1]{\mathrm{#1}}

\newcommand{\grSU}{\grp{SU}}
\newcommand{\grSO}{\grp{SO}}

\newcommand{\algSU}{\alg{su}}

\newcommand{\lrbrk}[1]{\left(#1\right)}
\newcommand{\bigbrk}[1]{\bigl(#1\bigr)}

\newcommand{\biggbrk}[1]{\biggl(#1\biggr)}

\newcommand{\lrsbrk}[1]{\left[#1\right]}
\newcommand{\bigsbrk}[1]{\bigl[#1\bigr]}
\newcommand{\Bigsbrk}[1]{\Bigl[#1\Bigr]}
\newcommand{\biggsbrk}[1]{\biggl[#1\biggr]}
\newcommand{\Biggsbrk}[1]{\Biggl[#1\Biggr]}
\newcommand{\ket}[1]{\mathopen{|}#1\mathclose{\rangle}}

\newcommand{\abs}[1]{{|#1|}}

\newcommand{\nn}{\nonumber}

\newcommand{\earel}[1]{\mathrel{}&\hspace{-2\arraycolsep}#1\hspace{-2\arraycolsep}&\mathrel{}}
\newcommand{\eq}{\earel{=}}

\def\[{\begin{equation}}
\def\]{\end{equation}}

\makeatletter
\def\mr@ignsp#1 {\ifx\:#1\@empty\else #1\expandafter\mr@ignsp\fi}%
\newcommand{\multiref}[1]{\begingroup
\xdef\mr@no@sparg{\expandafter\mr@ignsp#1 \: }%
\def\mr@comma{}%
\@for\mr@refs:=\mr@no@sparg\do{\mr@comma\def\mr@comma{,}\ref{\mr@refs}}%
\endgroup}
\makeatother

\newcommand{\hypref}[2]{\ifx\href\asklfhas #2\else\href{#1}{#2}\fi}

\newcommand{\secref}[1]{Sec.~\multiref{#1}}

\newcommand{\appref}[1]{App.~\multiref{#1}}

\newcommand{\figref}[1]{Fig.~\multiref{#1}}
\renewcommand{\eqref}[1]{(\multiref{#1})}

\ifx\href\asklfhas\newcommand{\href}[2]{#2}\fi

\newcommand{\comma}{\quad,\quad}
\newcommand{\unit}{\mathbbm{1}}

\newcommand{\Bsi}{\Upsilon}

\newcommand{\g}{\gamma}

\newcommand{\Smatrix}{\mathbbm{S}}  
\newcommand{\smatrix}{{S}}   
\newcommand{\Tmatrix}{\mathbbm{T}}  

\newcommand{\Asmatrix}{{A}}
\newcommand{\Bsmatrix}{{B}}
\newcommand{\Csmatrix}{{C}}
\newcommand{\Dsmatrix}{{D}}
\newcommand{\Esmatrix}{{E}}
\newcommand{\Fsmatrix}{{F}}
\newcommand{\Gsmatrix}{{G}}
\newcommand{\Hsmatrix}{{H}}
\newcommand{\Ksmatrix}{{K}}
\newcommand{\Lsmatrix}{{L}}

\newcommand{\lAA}{{a}}
\newcommand{\rAA}{{\dot{a}}}
\newcommand{\laa}{{\alpha}}
\newcommand{\raa}{{\dot{\alpha}}}
\newcommand{\lBB}{{b}}
\newcommand{\rBB}{{\dot{b}}}
\newcommand{\lbb}{{\beta}}
\newcommand{\rbb}{{\dot{\beta}}}
\newcommand{\lCC}{{c}}
\newcommand{\rCC}{{\dot{c}}}
\newcommand{\lcc}{{\gamma}}
\newcommand{\rcc}{{\dot{\gamma}}}
\newcommand{\lDD}{{d}}
\newcommand{\rDD}{{\dot{d}}}
\newcommand{\ldd}{{\delta}}
\newcommand{\rdd}{{\dot{\delta}}}

\newcommand{\levi}{\epsilon}
\newcommand{\energy}{\varepsilon}

\newcommand{\be}{\begin{eqnarray}}
\newcommand{\ee}{\end{eqnarray}}

\renewcommand{\vec}[1]{\mathbf{#1}}
\newcommand{\vecsigma}{\boldsymbol{\sigma}}
\newcommand{\ppp}{p_-,p_-'}

\begin{document}

\thispagestyle{empty}
\begin{flushright}\footnotesize
\texttt{hep-th/0704.3891}\\
\texttt{ITEP-TH-17/07}\\
\texttt{UUITP-07/07} \vspace{0.8cm}
\end{flushright}

\renewcommand{\thefootnote}{\fnsymbol{footnote}}
\setcounter{footnote}{0}

\begin{center}
{\Large\textbf{\mathversion{bold}
World-sheet scattering\\
in $AdS_5\times S^5$ at two loops
}\par}

\vspace{1.5cm}

\textrm{T.~Klose$^{1}$, T.~McLoughlin$^{2}$, J.~A.~Minahan$^{1}$ and K.~Zarembo$^{1}$\footnote{Also at ITEP, Moscow, Russia}} \vspace{8mm}

\textit{$^{1}$ Department of Theoretical Physics, Uppsala University\\
SE-751 08 Uppsala, Sweden}\\
\texttt{Thomas.Klose,Joseph.Minahan,Konstantin.Zarembo@teorfys.uu.se} \vspace{3mm}

\textit{$^{2}$ Department of Physics, The Pennsylvania State
University\\ University Park, PA 16802, USA}\\
\texttt{tmclough@phys.psu.edu} \vspace{3mm}


\par\vspace{1cm}

\textbf{Abstract} \vspace{5mm}

\begin{minipage}{14cm}
We study the $\AdS_5\times \Sphere^5$ sigma-model truncated to the near-flat-space
limit to two-loops in perturbation theory. In addition to extending previously known one-loop
results to the full $\grSU(2|2)^2$ S-matrix we calculate the
two-loop correction to the dispersion relation and then compute the complete two-loop S-matrix. The
result of the perturbative calculation can be compared with the appropriate limit of the conjectured
S-matrix for the full theory and complete agreement is found.
\end{minipage}

\end{center}

\vspace{0.5cm}

\newpage
\setcounter{page}{1}
\renewcommand{\thefootnote}{\arabic{footnote}}
\setcounter{footnote}{0}

\tableofcontents
\section{Introduction}

There has been much recent progress in the effort to completely
establish the AdS/CFT correspondence
\cite{Maldacena:1998re,Gubser:1998bc,Witten:1998qj}. The full
conjectured integrability of planar $\mathcal{N}=4$ Super Yang-Mills
\cite{Minahan:2002ve,Beisert:2003tq,Beisert:2003yb} and its dual
theory, the string sigma-model on an $\AdS_5\times\Sphere^5$ target
space \cite{Bena:2003wd}, has been instrumental in this progress. At
least for the question of gauge operators of infinite bare
dimension, 
computing the spectrum has basically come down to finding a
two-particle S-matrix \cite{Staudacher:2004tk} that can be
determined for both large and small values of the 't Hooft coupling.

For large 't Hooft coupling, the scattering is that of string
oscillators on the world-sheet, while for small 't Hooft coupling it
more closely resembles the scattering of magnons on a spin-chain.
Remarkably, as was shown by Beisert
\cite{Beisert:2005tm,Beisert:2006qh}, the S-matrix is almost
completely determined  by the underlying superalgebra
$\algSU(2|2)\times \algSU(2|2)$ with central extension, no matter
what the coupling. The only part of the S-matrix that cannot be
determined from the supergroup structure itself is an overall phase
factor (the dressing phase),
which was conjectured first in the form of an asymptotic series at
strong coupling \cite{Beisert:2006ib}, and then non-perturbatively
\cite{Beisert:2006ez}. First steps towards derivation of the
dressing phase from Bethe ansatz were taken in the recent work
\cite{Rej:2007vm,Sakai:2007rk}. The conjectured dressing phase makes
the S-matrix crossing-symmetric \cite{Janik:2006dc} and passes a
remarkable four-loop test at weak 't~Hooft coupling: decoration of
the Bethe equations with the conjectured phase modifies the
anomalous dimensions starting from four loops and such a
modification brings the Bethe-ansatz prediction for the cusp
anomalous dimension \cite{Beisert:2006ez} into agreement with the
explicit four-loop calculation \cite{Bern:2006ew,Cachazo:2006az}.

It is remarkable that explicit four-loop calculations in
$\mathcal{N}=4$ SYM are possible and it is certainly desirable to
reach comparable accuracy on the string side. Currently, state of
the art is the one-loop order: quantum corrections to the energies
of various classical string configurations have been computed in
\cite{Frolov:2002av,Frolov:2003tu,Gromov:2007cd}. The purpose of
this paper is to go beyond the one-loop order. Since the full
$\AdS_5\times\Sphere^5$ sigma-model \cite{Metsaev:1998it} is quite
complicated we make use of the simplifying limit proposed recently
by Maldacena and Swanson \cite{Maldacena:2006rv}.

 As in
\cite{Klose:2006zd,Klose:2007wq}, we will be interested in the
world-sheet S-matrix which can be directly compared to the
$\algSU(2|2)\times \algSU(2|2)$ S-matrix
\cite{Beisert:2005tm,Beisert:2006qh} with the conjectured dressing
phase \cite{Beisert:2006ib,Beisert:2006ez}.
The world-sheet S-matrix simplifies immensely in the
Maldacena-Swanson limit, but is nonetheless nontrivial since the
resulting sigma model is still interacting. The limit is taken by
scaling all momenta, such that $p\lambda^{1/4}$ is finite. The
momenta of the string excitations then sit in the ``near-flat''
region, between the noninteracting BMN regime
\cite{Berenstein:2002jq} and the classical giant magnon regime of
\cite{Hofman:2006xt}. For excitations in the near-flat region,
although there is no $\sin p/2$ factor in the dispersion relation
like in the case for giant magnons, the Lorentz invariance of the
BMN region is still broken by interaction terms. However, as we will
show this breaking of Lorentz invariance is rather mild, and in fact
can be restored if one compensates any Lorentz boost with a
rescaling of the world-sheet coupling constant.
It might be possible to argue that the S-matrix satisfies the usual
crossing symmetry as a consequence of the usual LSZ theorems, with
additional modifications due to this "soft" breaking of Lorentz
invariance. We shall see that the crossing symmetry is certainly
there at the level of Feynman diagrams.

The near-flat  limit also leads to a simplification of the Janik's
equation \cite{Janik:2006dc}. The odd solution will still be a sum
of dilogarithms, but the even phase simplifies tremendously and will
end up being the log of a rational function of the world-sheet
coupling (and so its contribution to the S-matrix is to multiply it
by a rational function).  It is simple to check that this function
is a solution to the near-flat limit of the BHL even equation.   The
S-matrix for the various processes also turns out to be a quadratic
polynomial of the world-sheet coupling multiplied by a common
function.

Computing quantum  corrections is much simpler in the near-flat
limit.  The quartic nature of the interaction terms makes the
computations similar to those found in $\phi^4$ theory in two
dimensions.  For two-point functions, supersymmetry prevents any
tadpoles from occuring, so there is no one-loop wave function
renormalization or mass-shift.  However, at the two-loop level 
there are sunset diagrams which induce radiative corrections to the
dispersion relation that agree with the predicted near-flat limit of
the dispersion relation in \cite{Maldacena:2006rv}.

We then consider  corrections to the four point amplitudes.  We will
compute these corrections up to the two-loop level, where we will
find agreement with  the near-flat limit of the BHL prediction. This
provides the first nontrivial check that goes beyond the tree level
AFS \cite{Arutyunov:2004vx} and one-loop HL \cite{Hernandez:2006tk}
dressing factor terms.
In carrying out these computations, we will see that the final
amplitudes for the different processes are very similar, as they
must be if they are to agree with the BHL S-matrix, but the road to
how these final amplitudes are reached can be significantly
different. For example, for certain $2\rightarrow 2$ bosonic
processes, there is a four-fermion interaction term that contributes
to the two-loop amplitude, while in other processes this interaction
term plays no role. In any case, the underlying supersymmetry must
play a crucial part in determining the final structure of these
amplitudes. In going from the amplitude to the S-matrix, we must
take into account the two-loop wave-function renormalization as well
as the two-loop mass-shift which will affect the Jacobian factor
that needs to be included.

This paper is  structured as follows.  In \secref{sec:nearflatmodel}
we review Maldacena and Swanson's action for the near-flat limit. In
\secref{sec:propagator} we consider the two-loop two-point
functions, where we explicitly compute the wave-function
renormalization and mass-shift. In \secref{sec:s-matrix} we derive
the near-flat limit of the conjectured S-matrix. In
\secref{sec:1loopamp} we find the one-loop four-point amplitudes
while in \secref{sec:2loopamp} we find these amplitudes at two
loops. In these last two sections we also show that these results
are in agreement with the results in \secref{sec:s-matrix}. In
\secref{sec:conclusion} we present our conclusions.  We also include
several appendices which contain some of the technical details of
our calculations.

\section{Near-flat-space model}
\label{sec:nearflatmodel}

Our starting point is the relatively simple light-cone action for
the reduced model of \cite{Maldacena:2006rv}(in the notation of
\cite{Klose:2007wq}):
\begin{eqnarray} \label{action}
 \mathcal{L}&=&\frac{1}{2}(\partial Y)^2-\frac{m^2}{2}\,Y^2
 +\frac{1}{2}(\partial Z)^2-\frac{m^2}{2}\,Z^2
 +\frac{i}{2}\,\psi \,\frac{\partial ^2+m^2}{\partial _-}\,\psi
 \nonumber \\[1mm]
 &&
 +\g\, (Y^2-Z^2)\bigbrk{(\partial _-Y)^2+(\partial _-Z)^2}
 +i\g\, (Y^2-Z^2)\psi\partial_-\psi
 \nonumber \\[2mm]
 &&
 +i\g\, \psi\bigbrk{\partial _-Y^{i'} \Gamma^{i'}
 +\partial_-Z^i\Gamma ^i}
 \bigbrk{Y^{i'} \Gamma^{i'}-Z^i\Gamma ^i}\psi
 \nonumber \\[1mm]
 &&
 -\frac{\g}{24}\bigbrk{\psi\Gamma^{i'j'}\psi\,\psi\Gamma^{i'j'}\psi
 -\psi\Gamma^{ij}\psi\,\psi\Gamma^{ij}\psi}.
\end{eqnarray}
The bosonic fields $Z^i$ and $Y^{i'}$ correspond to transverse excitations in
the $AdS_5$ and $S^5$ directions respectively and the fermions, $\psi$, are Majorana-Weyl
$\grSO(8)$ spinors of positive chirality.\footnote{See \appref{sec:indicies} for a more complete description
of the relevant conventions and notations.}  The action in \ref{action} is not invariant under world-sheet Lorentz transformations, but it is invariant under 8 independent linearly realized supersymmetries.

This action is the same as the near-flat space truncation of \cite{Maldacena:2006rv}, however as in
\cite{Klose:2007wq}, we have introduced the parameter $\g$ by rescaling the worldsheet coordinates
and furthermore we have integrated out the half of the original sixteen fermions which 
occured only quadratically in the action. The near-flat space action action of  \cite{Maldacena:2006rv} was obtained from $\AdS_5\times\Sphere^5$ string sigma-model by expanding about a  constant density solution boosted with rapidity $\sim \lambda^{1/4}$ in the $\sigma^-$ direction and so the above truncation should be equivalent to the full theory in the near-flat limit,
\[ \label{eqn:MSlimit}
p_- \sim \sqrt[4]{\lambda} \comma p_+ \sim \frac{1}{\sqrt[4]{\lambda}}\,
\]
provided we set
\[
\g=\frac{\pi}{\sqrt{\lambda}}
\]
and the mass, $m$, to be unity.

\section{Two-loop propagator}
\label{sec:propagator}

We now turn to the computation of the two-loop correction to the propagator. Firstly, we confirm that this leads to the expected mass shift and therefore the expected corrections to the dispersion relation. Secondly, for our two-loop scattering computation in \secref{sec:2loopamp}, it is necessary to know the residue of the pole in the propagator, which we determine here as well.

The dispersion relation in the original sigma model is expected to be
\[ \label{eqn:propagator-prediction}
\energy = m \sqrt{1 + \frac{1}{\g^2} \sin^2 \frac{\g p}{m}} \xrightarrow{\;\;\;\eqref{eqn:MSlimit}\;\;} \sqrt{m^2 + p^2 - \frac{\g^2 p_-^4}{3m^2}} \; .
\]
The second expression is the predicted exact dispersion relation in the near-flat limit \eqref{eqn:MSlimit}. We will now derive this dispersion relation from a Feynman diagram computation in the model \eqref{action}. This computation shows for the first time the emergence of the sine in the dispersion relation from the perturbation expansion of the string sigma-model.

The first correction to the propagator is of order $\g^2$ and the corresponding diagram is the sunset diagram drawn in \figref{fig:sunset} on page \pageref{fig:sunset}. Doing the combinatorics for the bosonic and the fermionic propagator, respectively, leads to
\[ \label{eqn:prop-combinatorics}
\begin{split}
\Amp_b(p) & =
32i\gamma^2 \lrsbrk{
2p_-^2 \lrbrk{2I_{110}+I_{200}}
-p_- \lrbrk{I_{111}+I_{210}}
+ \lrbrk{4I_{211}+I_{220}+I_{310}}
} \\
\Amp_f(p) & =
16i\gamma^2 \lrsbrk{
p_-^2 I_{100}
+ 2 p_- I_{200}
+ 6I_{111} + 14 I_{210} + I_{300}
}
\end{split}
\]
where
\[
I_{rst}(p) = \int \frac{d^2\vec{k}\,d^2\vec{q}}{(2\pi)^4} \frac{(k_-)^r \, (q_-)^s \, (p_- - k_- - q_-)^t}{(\vec{k}^2 - m^2)(\vec{q}^2 - m^2)[(\vec{p}-\vec{k}-\vec{q})^2 - m^2]} \; .
\]
This integral is the sunset diagram with $r$, $s$ and $t$ powers of the three momenta inserted into the numerator, cf.~\appref{app:sunset}. Some of these factors originate from derivative couplings, others are due to the extra power of $p_-$ in the fermionic propagator. We can simplify the expression for the amplitudes using the identity
\[
  p_- I_{rst} = I_{r+1,s,t} + I_{r,s+1,t} + I_{r,s,t+1} \; .
\]
Applying this identity repeatedly, we find that the amplitudes simplify to
\[ \label{eqn:prop-off-shell}
\Amp_b(p) = \frac{64}{3}i\gamma^2 p_-^4 I_{000}(\vec{p}^2)
\comma
\Amp_f(p) = \frac{64}{3}i\gamma^2 p_-^3 I_{000}(\vec{p}^2)
\; ,
\]
where $I_{000}$ is a function of $\vec{p}^2$ only. It is interesting to see how the very different structures in \eqref{eqn:prop-combinatorics} reduce to essentially the same expression. We perform this integral in \appref{app:sunset} and find for the on-shell amplitudes
\[
\Amp_b(p) = i\gamma^2 \frac{p_-^4}{3 m^2}
\comma
\Amp_f(p) = i\gamma^2 \frac{p_-^3}{3 m^2}
\; .
\]

In order to find the corrected dispersion relation, we consider the iteration of sunset diagrams \eqref{eqn:prop-off-shell}. Via a geometric series this leads to the corrected propagator
\[ \label{eqn:def-disp-renorm}
\frac{i}{\vec{p}^2 - m^2 + \frac{64}{3}i\gamma^2 p_-^4 I_{000}(\vec{p})} \stackrel{!}{=} \frac{i Z(p_-)}{2p_+ - \Sigma(p_-)} + \mbox{finite as $2p_+\to\Sigma(p_-)$} \; ,
\]
where there is an extra factor of $p_-$ in the numerator for the fermionic propagator. The right hand side of \eqref{eqn:def-disp-renorm} defines the position $\Sigma(p_-)$ and the residue $Z(p_-)$ of the pole in the propagator in the $2p_+$ plane. Note that for our definition of the light-cone momenta \eqref{eqn:lc-momenta}, $2p_+$ is the appropriate ``energy'' for time evolution in $\sigma^+$ direction.

The dispersion relation is determined by the pole in the propagator. To order $\g^2$ we only need the on-shell value \eqref{eqn:I000-on-shell} of the integral $I_{000}$ and find
\[ \label{dispersion}
 p_+(p_-) = \Half \Sigma(p_-) = \frac{m^2}{4p_-} - \frac{\g^2 p_-^3}{12m^2} \; .
\]
Using $\energy^2 - p^2 = 4 p_+ p_-$, we convert this equation into the form $\energy(p)$ and find that this dispersion relation exactly agrees with the prediction in \eqref{eqn:propagator-prediction}.

For computing the residue we also need the on-shell value of the first derivative of $I_{000}$ with respect to $\vec{p}^2$. Taking this integral from \eqref{eqn:I000prime}, we find the wave-function renormalization to order $\g^2$ to be
\[\label{wfren}
 Z(p_-) = \frac{1}{2p_-} \lrsbrk{
 1 - \frac{\g^2}{m^4} \lrbrk{ \frac{1}{\pi^2} - \frac{1}{12} } p_-^4
 } \; .
\]
This correction is an important contribution to the two-loop amplitudes which we compute in \secref{sec:2loopamp}. It will turn out that this correction cancels the entire wineglass contribution in the $t$-channel.

\section{$\grSU(2|2)$ S-matrix}
\label{sec:s-matrix}

The $\grSU(2|2)$ scattering matrix is expressed in terms of the
following kinematic variables\footnote{We use the string
normalization of momenta, which differs from the spin chain
normalization in \cite{Beisert:2005tm} by a factor of $2\pi
/\sqrt{\lambda }$.}:
\begin{equation}\label{xpm}
x_{\pm}(p) = \frac{1+\sqrt{1+P^2}}{P}\,\,{\rm e}\,^{\pm\frac{i\pi
p}{\sqrt{\lambda }}}\;,\qquad P=\frac{\sqrt{\lambda }}{\pi
}\,\sin\frac{\pi p}{\sqrt{\lambda }}\;.
\end{equation}
For the S-matrix components, we use the conventions of \cite{Klose:2006zd}:
\begin{align}
\quad
\smatrix_{\lAA\lBB}^{\lCC\lDD} & = \Asmatrix \,\delta_\lAA^\lCC \delta_\lBB^\lDD
                                 + \Bsmatrix \,\delta_\lAA^\lDD \delta_\lBB^\lCC \; , && \qquad\quad &
\smatrix_{\lAA\lBB}^{\lcc\ldd} & = \Csmatrix \,\levi_{\lAA\lBB} \levi^{\lcc\ldd} \; , && \quad \nn \\
\smatrix_{\laa\lbb}^{\lcc\ldd} & = \Dsmatrix \,\delta_\laa^\lcc \delta_\lbb^\ldd
                                 + \Esmatrix \,\delta_\laa^\ldd \delta_\lbb^\lcc \; , &&  &
\smatrix_{\laa\lbb}^{\lCC\lDD} & = \Fsmatrix \,\levi_{\laa\lbb} \levi^{\lCC\lDD} \; , &&  \label{smat0} \\
\smatrix_{\lAA\lbb}^{\lCC\ldd} & = \Gsmatrix \,\delta_\lAA^\lCC \delta_\lbb^\ldd \; , &&  &
\smatrix_{\lAA\lbb}^{\lcc\lDD} & = \Hsmatrix \,\delta_\lAA^\lDD \delta_\lbb^\lcc \; , && \nn \\
\smatrix_{\laa\lBB}^{\lcc\lDD} & = \Lsmatrix \,\delta_\laa^\lcc \delta_\lBB^\lDD \; , &&  & \smatrix_{\laa\lBB}^{\lCC\ldd} & = \Ksmatrix \,\delta_\laa^\ldd \delta_\lBB^\lCC \; . && \nn
\end{align}
The explicit expressions for matrix elements are
\cite{Beisert:2005tm}%
\footnote{Comparison with the explicit tree-level calculations
\cite{Klose:2006zd} shows that the scattering in the sigma-model is
described by the $\grSU(2|2)$ S-matrix in its canonical form
\cite{Arutyunov:2006yd} and should include phase factors $e^{\pm \pi
ip_1/\sqrt{\lambda}}$ and $e^{\pm \pi ip'_1/\sqrt{\lambda }}$ that
multiply the S-matrix elements in particular combinations. In other
possible forms, which are related to the canonical S-matrix by
state-dependent unitary transformations \cite{Arutyunov:2006yd},
$e^{\pm \pi ip_1/\sqrt{\lambda }}$, $e^{\pm \pi ip'_1/\sqrt{\lambda
}}$
 are replaced by arbitrary functions of $p$, $p'$ \cite{Beisert:2006qh}
 (for instance by $1$ as in the original proposal \cite{Beisert:2005tm}).
 It is interesting to note that in the near-flat-space limit the phase
 factors scale away and can be dropped altogether.}:
\begin{eqnarray}\label{sma}
 &&\Asmatrix=\frac{x'_--x_-}{x'_--x_+}\,\,
 \frac{1-\frac{1}{x'_-x_+}}{1-\frac{1}{x'_+x_+}}
 \; , \nonumber \\
 &&\Bsmatrix=\frac{x'_+-x_-}{x'_--x_+}\left(
 1-\frac{x'_--x_-}{x'_+-x_-}\,\,
 \frac{1-\frac{1}{x'_-x_+}}{1-\frac{1}{x'_+x_+}}
 \right)
  \; , \nonumber \\
 &&\Csmatrix=\frac{i\eta \eta '}{x_+x'_+}\,\,
 \frac{1}{1-\frac{1}{x'_+x_+}}\,\,
 \frac{x'_--x_-}{x'_--x_+}\,\,{\rm e}\,^{\frac{i\pi p'_1}{\sqrt{\lambda }}}
 \; , \nonumber \\
 &&\Dsmatrix=\frac{x'_+-x_+}{x'_--x_+}\,\,
 \frac{1-\frac{1}{x'_+x_-}}{1-\frac{1}{x'_-x_-}}\,\,{\rm e}\,^{\frac{i\pi (p'_1-p_1)}
 {\sqrt{\lambda }}}
 \; , \nonumber \\
 &&\Esmatrix = 1-\frac{x'_+-x_+}{x'_--x_+}\,\,
 \frac{1-\frac{1}{x'_+x_-}}{1-\frac{1}{x'_-x_-}}\,\,{\rm e}\,^{\frac{i\pi (p'_1-p_1)}
 {\sqrt{\lambda }}}
 \; , \nonumber \\
 &&\Fsmatrix=-\frac{i(x_+-x_-)(x'_+-x'_-)}{\eta \eta
 'x_-x'_-}\,\,
 \frac{1}{1-\frac{1}{x'_-x_-}}\,\,
 \frac{x'_+-x_+}{x'_--x_+}\,\,{\rm e}\,^{-\frac{i\pi p_1}{\sqrt{\lambda }}}
 \; , \nonumber \\
 &&\Gsmatrix=\frac{x'_+-x_+}{x'_--x_+}\,\,{\rm e}\,^{-\frac{i\pi p_1}{\sqrt{\lambda }}}
  \; , \qquad \quad
 \Hsmatrix=\frac{\eta }{\eta '}\,\,\frac{x'_+-x'_-}{x'_--x_+}\,\,{\rm e}\,^{\frac{i\pi (p'_1-p_1)}
 {\sqrt{\lambda }}}
 \; , \nonumber \\
 &&\Lsmatrix=\frac{x'_--x_-}{x'_--x_+}\,\,{\rm e}\,^{\frac{i\pi p'_1}{\sqrt{\lambda }}}
 \; , \qquad \qquad
 \Ksmatrix=\frac{\eta' }{\eta}\,\,\frac{x_+-x_-}{x'_--x_+}
 \; ,
\end{eqnarray}
where $x_\pm\equiv x_\pm(p)$, $x'_\pm\equiv x_\pm(p')$ and
\[
\eta=\abs{x_--x_+}^{1/2},\qquad \eta '=\abs{x'_--x'_+}^{1/2} \; .
\]

The sigma-model scattering matrix is the tensor product of the two
$\grSU(2|2)$ S-matrices. The world-sheet scattering amplitudes are
thus quadratic in the $A,B,C,D,\ldots $. In addition the world-sheet
scattering matrix contains an overall phase factor:
\begin{equation}\label{amp}
 \Smatrix=\frac{1-\frac{1}{x'_+x_-}}{1-\frac{1}{x'_-x_+}}\,\,
 \frac{x'_--x_+}{x'_+-x_-}\,
 \,{\rm e}\,^{i\theta (p,p')}\,
 \,\smatrix\otimes\smatrix \; ,
\end{equation}
where $\theta (p,p')$ is the dressing phase. For reader's
convenience we have written the action of $\Smatrix$ on all
two-particle states in \appref{app:everything} in order to see which
matrix elements govern which processes.

The dressing phase has the following general form
\cite{Arutyunov:2004vx,Beisert:2005wv}:
\begin{equation}\label{chidef}
 \theta(p,p') = \sum_{r,s=\pm} r s \,\chi(x_r,x'_s)
 \; .
\end{equation}
The function $\chi(x,y)$ is anti-symmetric in $x$ and $y$ and can
be expanded in asymptotic power series in $\pi /\sqrt{\lambda }$. We
only need the first three orders of this expansion:
\begin{eqnarray}\label{chick}
 \chi (x,y)&=&\frac{\sqrt{\lambda }}{2\pi }\left(x-y\right)
 \left(1-\frac{1}{xy}\right)\ln\left(1-\frac{1}{xy}\right)
 \nonumber \\
 &&+\int_{0}^{1}\frac{dt}{\pi t}\,\,\ln\left[
 \frac{(1-t)^2xy-(t-x)^2}{(1-t)^2xy-(t+x)^2}
 \,\,
 \frac{(1-t)^2xy-(t+y)^2}{(1-t)^2xy-(t-y)^2}
 \right]
 \nonumber \\
 &&+\frac{\pi }{3\sqrt{\lambda }}\,\,
 \frac{xy+1}{xy-1}\,\,\frac{x-y}{\left(x^2-1\right)\left(y^2-1\right)}+\ldots \; .
\end{eqnarray}
The first line is the AFS tree-level phase \cite{Arutyunov:2004vx},
the second line is the HL one-loop correction
\cite{Hernandez:2006tk} and the third line is taken from
\cite{Beisert:2006ib}. The integral in the one-loop phase can be
expressed in terms of the dilogarithms, but for our purposes the
integral representation is more convenient.  The first and last lines are part of BHL's even phase, while the middle line makes up the entire odd phase.

In the near-flat limit, the kinematic variables $x_\pm$ approach
$-1$. However, the S-matrix contains many expressions of the form
$x_rx'_s-1$ or $x_r-x'_s$ which vanish at $x_r=-1=x'_s$. Plugging in
$-1$ for $x_\pm$, $x'_\pm$ produces singularities and we need to
keep the next term in the expansion:
\begin{equation}\label{exex}
 x_\pm=-1-\frac{1}{p_-}\pm\frac{i\pi }{\sqrt{\lambda }}\,p_-+\ldots \; .
\end{equation}
The second and the third terms are small compared to one (they are
of order $\order(\lambda ^{-1/4})$) and should be omitted wherever
$-1$ does not cancel.

We thus get
\begin{equation}\label{Smtr}
 \Smatrix=\frac{1-\frac{i\pi }{\sqrt{\lambda }}\,p_-p'_-\,\frac{p'_--p_-}{p'_-+p_-}}
 {1+\frac{i\pi }{\sqrt{\lambda }}\,p_-p'_-\,\frac{p'_--p_-}{p'_-+p_-}}\,\,
 \frac{\,{\rm e}\,^{i\theta (p,p')}}
 {1+\frac{\pi ^2}{\lambda }\,p_-^2p'{}^2_-\left(\frac{p'_-+p_-}{p'_--p_-}\right)^2}
 \,\smatrix\otimes\smatrix
\end{equation}
where the matrix elements are as in (\ref{smat0}) with\footnote{We
chose to pull out a common factor of $\left(1-\frac{i\pi
}{\sqrt{\lambda }}\,p_-p'_-\,\frac{p'_-+p_-}{p'_--p_-}\right)^{-1}$
from $\smatrix$.}
\begin{align} \label{limmatel}
 \Asmatrix & =1+\frac{i\pi}{\sqrt{\lambda}}\,p_-p'_-\,\frac{p'_--p_-}{p'_-+p_-} \; , &
 \Bsmatrix & =-\Esmatrix=\frac{4i\pi }{\sqrt{\lambda}}\,\,\frac{p_-^2p'{}^2_-}{p_-'^2-p_-^2} \; , \nn \\
 \Dsmatrix & =1-\frac{i\pi }{\sqrt{\lambda}}\,p_-p'_-\,\frac{p'_--p_-}{p'_-+p_-} \; , &
 \Csmatrix & =\Fsmatrix=\frac{2i\pi }{\sqrt{\lambda }}\,\,\frac{p_-^{3/2} p_-'^{3/2}}{p_-'+p_-} \; , \nn \\
 \Gsmatrix & =1+\frac{i\pi }{\sqrt{\lambda}}\,p_-p'_- \; , &
 \Hsmatrix & =\Ksmatrix=\frac{2i\pi}{\sqrt{\lambda }}\,\,\frac{p_-^{3/2} p_-'^{3/2}}{p_-'-p_-} \; , \nn \\
 \Lsmatrix & =1-\frac{i\pi }{\sqrt{\lambda}}\,p_-p'_- \; . &
\end{align}
We should stress that these expressions are exact in the
near-flat limit. For comparison to the two-loop calculation in
\secref{sec:2loopamp} we need to further expand in $\pi
p_-^2/\sqrt{\lambda }$.

When expanding the phase in $\pi /\sqrt{\lambda }$ it is important
to remember that it implicitly depends on $\lambda $ through
$x_\pm$, apart from the explicit dependence manifest in
(\ref{chick}). In particular the tree-level term in (\ref{chick})
contains a two-loop correction to the phase. The substitution of
(\ref{exex}) into (\ref{chidef}), (\ref{chick}) yields after a
lengthy but straightforward calculation:
\begin{eqnarray}
 \theta (p,p')&=&\frac{2\pi }{\sqrt{\lambda }}\,p_-p'_-\,
 \frac{p'_--p_-}{p'_-+p_-}
 -\frac{4\pi ^3}{3\lambda^{3/2}}\,p_-^3p'{}^3_-\,
 \frac{(p'_--p_-)(p'{}^2_-+p_-p'_-+p^2_-)}{(p'_-+p_-)^3}
 \nonumber \\
 &&+\frac{8\pi }{\lambda }\,\,\frac{p^3_-p'{}^3_-}{p'{}^2_--p^2_-}
 \left(1-\frac{p'{}^2_-+p_-^2}{p'{}^2_--p_-^2}\,\ln\frac{p'_-}{p_-}\right)
 \nonumber \\
 &&+\frac{2\pi ^3}{3\lambda ^{3/2}}\,p^3_-p'{}^3_-\,
 \frac{(p'_--p_-)(p'{}^2_-+4p_-p'_-+p^2_-)}{(p'_-+p_-)^3}+\ldots \,.
\end{eqnarray}
Omitting $\order\lrbrk{\bigbrk{\frac{p}{\sqrt[4]{\lambda}}}^6}$ terms this can be written in the
following nice form, suggested by the main scattering term,
\begin{equation}\label{nfphase}
 \theta (p,p')=-i\ln
 \frac{1+\frac{i\pi }{\sqrt{\lambda}}\,p_-p'_-\,\frac{p'_--p_-}{p'_-+p_-}}
 {1-\frac{i\pi }{\sqrt{\lambda }}\,p_-p'_-\,\frac{p'_--p_-}{p'_-+p_-}}
 +\frac{8\pi }{\lambda }\,\,\frac{p^3_-p'{}^3_-}{p'{}^2_--p^2_-}
 \left(1-\frac{p'{}^2_-+p_-^2}{p'{}^2_--p_-^2}\,\ln\frac{p'_-}{p_-}\right),
\end{equation}
where the first (second) term comes from the even (odd) phase
\footnote{ We suspect that the even part of the phase in  \ref{nfphase} is
valid to all orders in $\gamma $. We checked this by taking the
near-flat limit of the BHL phase to order $\gamma ^{11}$.  Furthermore, one can readily see that it solves the near flat limit of the even crossing relation of (2.13) in \cite{Beisert:2006ib} . It should
be possible to prove (or disprove) this fact by inspecting the 
integral representation of the phase found in \cite{Dorey:2007xn}. }.
Equation (\ref{Smtr}) becomes
\begin{equation}\label{smatttr}
 \Smatrix = S_0\,\smatrix\otimes\smatrix
 \quad\mbox{with}\quad
 S_0 = \frac{\,\,e^{
 \frac{8\pi i}{\lambda }\,\,\frac{p^3_-p'{}^3_-}{p'{}^2_--p^2_-}
 \left(1-\frac{p'{}^2_-+p_-^2}{p'{}^2_--p_-^2}\,\ln\frac{p'_-}{p_-}\right)
 }}
 {1+\frac{\pi ^2}{\lambda }\,p_-^2p'{}^2_-\left(\frac{p'_-+p_-}{p'_--p_-}\right)^2}
\end{equation}
where $\smatrix$ is given by (\ref{smat0}), (\ref{limmatel}). At the
end, the dressing phase almost completely cancels the main
scattering phase, and the two-loop prediction for the scattering
amplitude turns out to be rather compact. We should stress that
(\ref{smatttr}) is only accurate up to $\order(1/\lambda ^2)$ (the full expression is expected to contain dilogarithms from the odd phase)  while
matrix elements (\ref{limmatel}) are exact in the near-flat
sigma-model.

In order to facilitate the comparison with the results from the
world-sheet computation, let us discuss the first few orders of
\eqref{smatttr}. The $n$-th loop contribution to the two-particle
S-matrix is of order $\g^{n+1} =
\bigbrk{\frac{\pi}{\sqrt{\lambda}}}^{n+1}$ and we denote it by
$\Smatrix^{(n)}$. Now, we observe that the prefactor $S_0$ in
\eqref{smatttr} does not have a term of order $\g =
\frac{\pi}{\sqrt{\lambda}}$ and that the coefficients in
\eqref{limmatel} stop at order $\g = \frac{\pi}{\sqrt{\lambda}}$.
Hence, the tree-level contribution to the S-matrix $\Smatrix^{(0)}$
originates only from the matrix elements in \eqref{limmatel}, the
one-loop contribution $\Smatrix^{(1)}$ receives additional terms
from the prefactor $S_0$ and the two-loop contribution is of the
form
\[ \label{smatttr2}
 \Smatrix^{(2)} = \frac{\pi^2}{\lambda} \lrsbrk{
 - p_-^2 p_-'^2 \lrbrk{\frac{p_-'+p_-}{p_-'-p_-}}^2
 + \frac{8i}{\pi} \, \frac{p^3_-p_-'^3}{p_-'^2-p_-^2}
   \lrbrk{1-\frac{p_-'^2+p_-^2}{p_-'^2-p_-^2}\,\ln\frac{p'_-}{p_-}}
 } \Smatrix^{(0)} \; ,
\]
i.e. the two-loop piece reproduces the tree-level S-matrix multiplied by a factor that is universal for all scattering processes.

We close this section by noting that the S-matrix can be put into a  form that looks almost relativistic.  
Under boosts the momenta, derivatives and fields transform as
\[
  p_\pm \to \alpha^{\mp1} p_\pm
  \comma
  \partial_\pm \to \alpha^{\mp1} \partial_\pm
  \comma
  Z,Y \to Z,Y
  \comma
  \psi \to \sqrt{\alpha} \, \psi \; ,
\]
where $\alpha$ is the boost parameter. If these transformation are accompanied by a rescaling of the coupling $\g \to \alpha^{-2} \, \g$, then the Lagrangian \eqref{action} is invariant under these transformations. As a consequence the S-matrix can be written as a function of a momentum dependent, but boost invariant coupling
\[
 \tilde{\g} = \tilde{\g}(p,p') = \g \, p_- p_-'
\]
and the relative rapidity $\theta = \arcsinh\frac{p}{m} - \arcsinh\frac{p'}{m} = \ln \frac{p_-'}{p_-}$.
Rewriting \eqref{smatttr} and \eqref{limmatel}, we find
\begin{equation}
 \Smatrix=
 \frac{\,\,e^{\mbox{$
 \frac{4i\tilde{\g}^2}{\pi} \, \frac{1-\theta \coth\theta}{\sinh\theta}$}}}
 {1+\tilde{\g}^2\,\coth^2\frac{\theta}{2}}
 \,\smatrix\otimes\smatrix \, ,
\end{equation}
with
\begin{align}
 \Asmatrix & =1+i\tilde{\g} \tanh\tfrac{\theta}{2} \, , &
 \Bsmatrix & =-\Esmatrix= 2i\tilde{\g} \csch\tfrac{\theta}{2} \, , \nn \\
 \Dsmatrix & =1-i\tilde{\g} \tanh\tfrac{\theta}{2} \, , &
 \Csmatrix & =\Fsmatrix=i\tilde{\g} \sech\tfrac{\theta}{2} \, , \nn \\
 \Gsmatrix & =1+i\tilde{\g} \, , &
 \Hsmatrix & =\Ksmatrix=i\tilde{\g} \csch\tfrac{\theta}{2} \, , \nn \\
 \Lsmatrix & =1-i\tilde{\g} \, . &
\end{align}
It would be interesting to see if a proof of crossing symmetry can be obtained given this relatively mild breaking of the Lorentz invariance.

\section{One-loop amplitudes}
\label{sec:1loopamp}

In this section we present the general bosonic one-loop amplitudes and S-matrices for $2\rightarrow2$ magnon scattering in the near-flat limit. These results generalize the case of $ZY\rightarrow ZY$ presented in \cite{Klose:2007wq}.

\begin{figure}
\begin{center}
\subfigure[s-channel]{\includegraphics[scale=0.6]{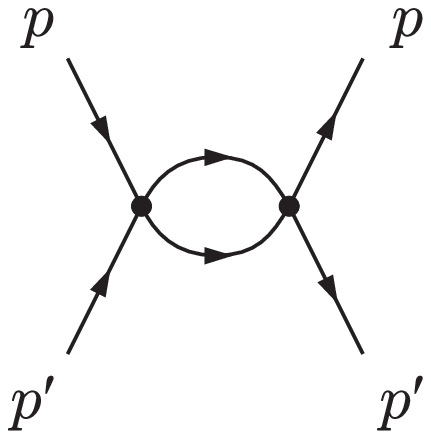}} \hspace{12mm}
\subfigure[t-channel]{\includegraphics[scale=0.6]{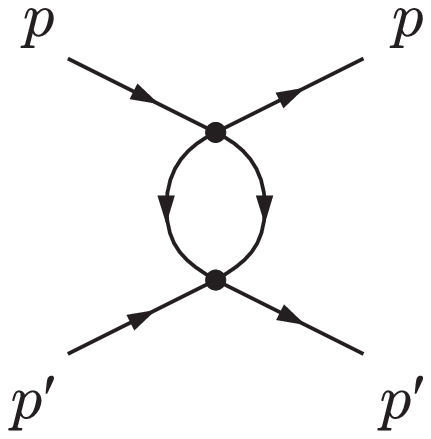}} \hspace{12mm}
\subfigure[u-channel]{\includegraphics[scale=0.6]{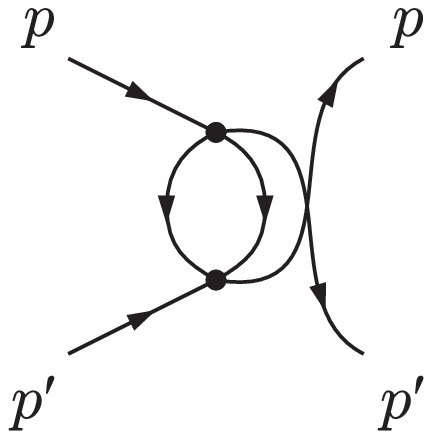}}
\caption{\textbf{One-loop graphs.} The basic structure for $s$, $t$ and $u$ channel graphs.}
\label{fig:oneloopgraphs}
\end{center}
\end{figure}

For all processes, there are three basic diagrams which are shown in \figref{fig:oneloopgraphs}.  We call these graphs the $s$, $t$ and $u$-channel graphs. Within each of these graphs, there can be several contributions to the complete loop in that channel. However, summing over the contributions will lead to three basic structures for the one-loop amplitudes.  The first of these is a structure associated with forward scattering, the second is a permutation structure and the third is a trace like structure. The latter two structures are related to each other through crossing symmetry.

The one-loop amplitudes are relatively straightforward to carry out.  For an amplitude of forward scattering type (for example $Z_1(p)Y_1(p')\longrightarrow Z_1(p)Y_1(p')$), the amplitude is found to be
\[ \label{1loopforward}
\begin{split}
\Amp^{(1)}_\mathrm{forward}(p_-,p_-') = -8\g^2 \Bigsbrk{ &  (p_-'+p_-)^2(p_-'^2 + p_-^2) I_{00}(p,-p') -8p_-^2 p_-'^2 I_{00}(p,p) \\
& + (p_-'-p_-)^2 (p_-'^2+p_-^2) I_{00}(p,p')} \; ,
\end{split}
\]
where $I_{00}(p,p')$ is the $u$-channel loop integral defined in \eqref{oneloopbubble}.  The $s$ and $t$ channel integrals are given by analytically continuing $p_-'$ to $-p_-'$ and letting $p_-'\to p_-$, respectively.
Substituting the results for the integrals into \eqref{1loopforward} gives
\[ \label{1loopffin}
\begin{split}
\Amp^{(1)}_\mathrm{forward}(p_-,p_-') =\ & 16i\g^2 \, \frac{p_-^2p_-'^2}{\pi(p_-'^2-p_-^2)} \lrsbrk{1-\frac{p_-'^2+p_-^2}{p_-'^2-p_-^2}\ln\frac{p_-'}{p_-}} \\
& -4\g^2\,\frac{p_-p_-'(p_-'+p_-)(p_-^2+p_-'^2)}{p_-'-p_-} \; .
\end{split}
\]
This result was previously derived in \cite{Klose:2007wq}.

The next type of scattering process is of the permutation type, where the outgoing $p$ and $p'$ are exchanged with a forward scattering process. In this case, summing over the contributions to the Feynman diagrams, we find
\[\label{1loopperm}
\Amp^{(1)}_\mathrm{perm}(p_-,p_-') = -16\g^2 \Bigsbrk{p_-p_-'(p_-'+p_-)^2I_{00}(p,-p') + p_-p_-'(p_-'+p_-)^2I_{00}(p,p')} \; ,
\]
where in these processes the contribution to the $t$-channel cancels out and the $u$-channel integral comes with the same kinematic factor as the $s$-channel. Substituting for the integrals into \eqref{1loopperm} we arrive at
\[\label{1looppfin}
\Amp^{(1)}_\mathrm{perm}(p_-,p_-') = -8\g^2\,\frac{p_-^2p_-'^2(p_-'+p_-)}{p_-'-p_-} \; .
\]

Finally the processes of trace type, which are of the form $A\bar A\longrightarrow B\bar B$, where $A$ and $B$ are any one of the fields and $\bar A$ and $\bar B$ are there conjugates is given by
\[\label{1looptrace}
\Amp^{(1)}_\mathrm{trace}(p_-,p_-') = 16\g^2 \Bigsbrk{p_-p_-'(p_-'-p_-)^2I_{00}(p,-p') + p_-p_-'(p_-'-p_-)^2I_{00}(p,p')}\; ,
\]
which after substituting for the integrals gives
\[\label{1looptfin}
\Amp^{(1)}_\mathrm{trace}(p_-,p_-') = 8\g^2\,\frac{p_-^2p_-'^2(p_-'-p_-)}{p_-'+p_-} \; .
\]
The amplitudes in \eqref{1looppfin} and \eqref{1looptfin} are related by crossing symmetry by taking $p_-'\to -p_-'$.  However, there is a subtlety in the analytic continuation, since the amplitudes were obtained by continuing around a log cut.  When continuing, say, $\Amp^{(1)}_\mathrm{trace}(p_-,p_-')$
to $\Amp^{(1)}_\mathrm{trace}(p_-,-p_-')$, one continues onto a different branch, hence leading to an extra minus sign.

\section{Two-loop amplitudes}
\label{sec:2loopamp}

In this section we compute the two-loop amplitudes for various four-point processes and show that there is complete agreement with the S-matrix results in \secref{sec:s-matrix}. One consequence of the structure of the S-matrix is that the two-loop amplitudes should be related to the tree amplitudes by a universal factor $\g^2F(p_-,p_-')$, which we will explicitly show.

In order to obtain the S-matrix, one must take into account the wave-function renormalization of the external legs as well as a Jacobian factor that arises when converting $\delta$-functions for overall conservation of energy and momentum to $\delta$-functions for individual momenta.  Moreover, there is a two-loop contribution to this Jacobian due to the two-loop mass-shift.  The contributions from the wave-function renormalization and Jacobian will cancel off against certain terms in the amplitude to give very compact expressions for the S-matrix.

\begin{figure}
\begin{center}
\subfigure[]{\includegraphics[scale=0.6]{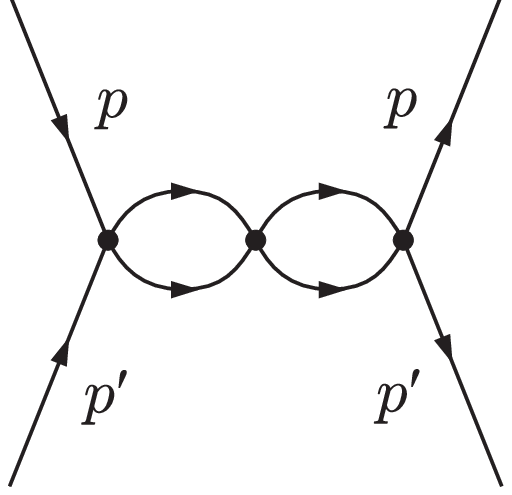}} \hspace{8mm}
\subfigure[]{\includegraphics[scale=0.6]{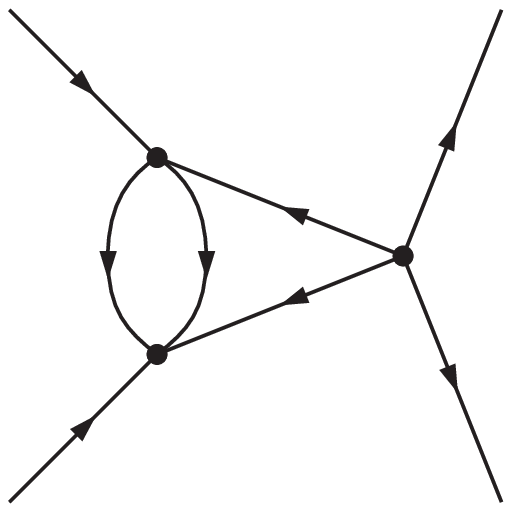}} \hspace{8mm}
\subfigure[]{\includegraphics[scale=0.6]{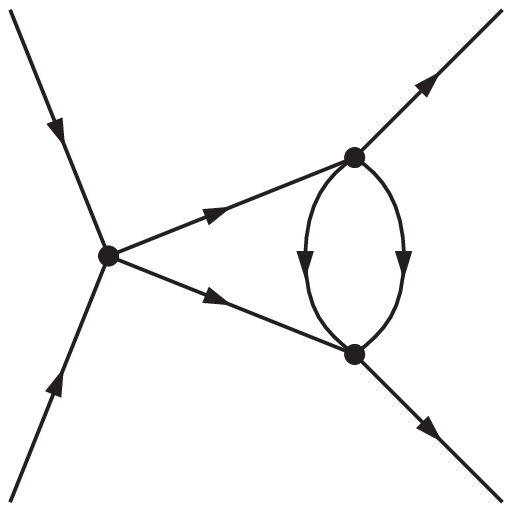}} \\
\subfigure[]{\includegraphics[scale=0.6]{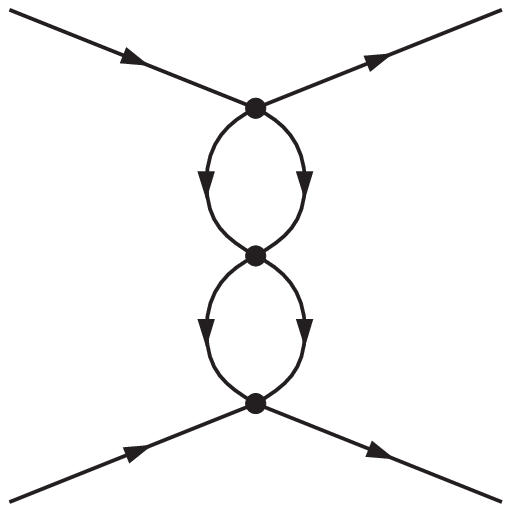}} \hspace{8mm}
\subfigure[]{\includegraphics[scale=0.6]{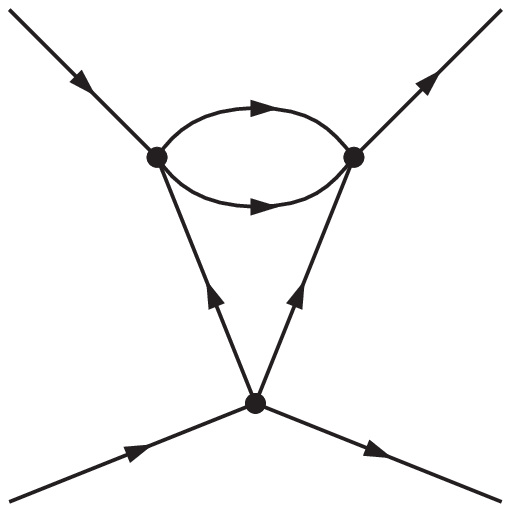}} \hspace{8mm}
\subfigure[]{\includegraphics[scale=0.6]{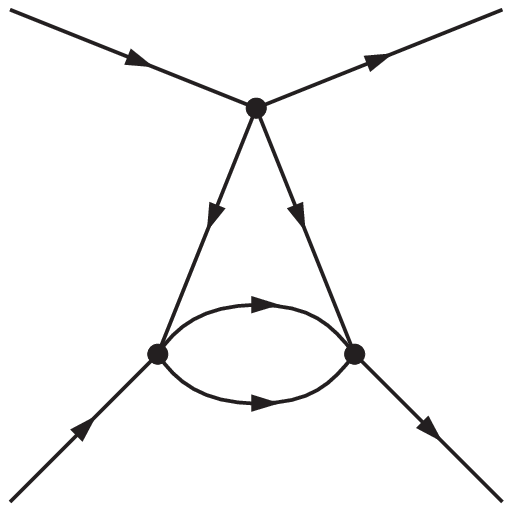}} \\
\subfigure[]{\includegraphics[scale=0.6]{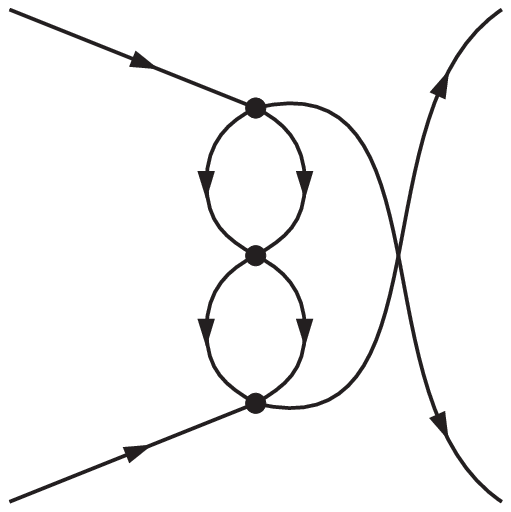}} \hspace{8mm}
\subfigure[]{\includegraphics[scale=0.6]{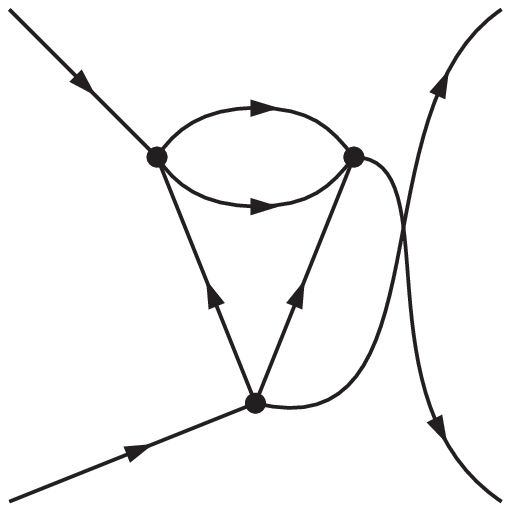} \hspace{8mm} \label{fig:twoloopgraphs:uwine}}
\subfigure[]{\includegraphics[scale=0.6]{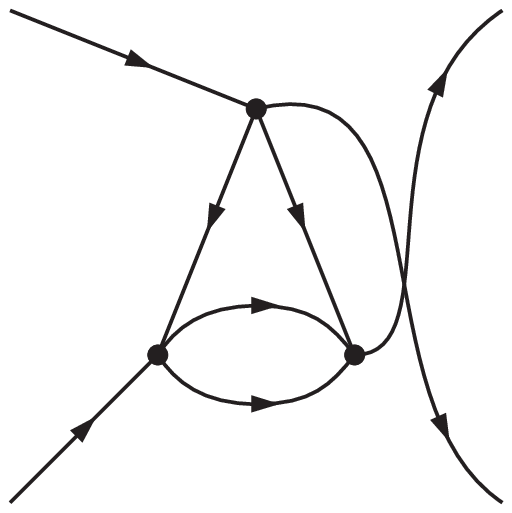}}
\caption{\textbf{Two-loop graphs.} The first line shows the $s$-channel diagrams, the second line the $t$-channel and the last $u$-channel. To the diagrams in the first column, we refer to as ``double bubble'', to the second column as ``wineglass'' and to the third as ``inverse wineglass''.}
\label{fig:twoloopgraphs}
\end{center}
\end{figure}

Since all interaction terms in (\ref{action}) are four-point, the general structure for the
two-loop Feynman diagrams have the form shown in \figref{fig:twoloopgraphs}.  The diagrams fall into the 3 general classes, ``double bubble'', ``wineglass'' and ``inverse wineglass'' for each of the $s$, $t$ and $u$ channels.  The bosonic vertices all come with two powers of $p_-$, a vertex with two bosons and two fermions has one power of $p_-$, while the four-fermion vertex has no powers of momenta.  The fermion propagator also comes with a factor of $p_-$, therefore the amplitudes  will all have world-sheet spin $-6$.  Naive power counting might indicate that these diagrams are divergent, however the two-dimensional Lorentz invariance of the free theory insures that these divergences are not there.

Let us start with the easiest set of diagrams to evaluate, the $t$-channel double bubble.  For these diagrams,  no external momentum flows through the internal propagators.  One can argue that there must be at least two powers of the internal momenta in the numerators of the two-loop integrals, which by the Lorentz invariance of the free theory,  must be zero, and so the $t$-channel bubbles all have $\Amp^\mathrm{bubble}_t(p,p')=0$.

The next set of diagrams we consider are the $u$ and $s$ channel double bubbles. Different processes have different combinatoric factors contributing to the loop integrals, but their final results all reduce to the same form, with the $u$-channel given by
\[\label{ububble}
\Amp^\mathrm{bubble}_u(p,p') = \g^2  \Amp^{(0)}(p_-,p_-') \, F^\mathrm{bubble}(p_-,p_-') \; ,
\]
where $\Amp^{(0)}(p_-,p_-')$ is the tree-level amplitude for the corresponding process and
\[
F^\mathrm{bubble}(p_-,p_-') = - \bigsbrk{ 2(p'_--p_-)^2 I_{00}(p_-,p_-') }^2 \; .
\]
$I_{rs}(p_-,p_-')$ is the one-loop $u$-channel integral defined in (\ref{oneloopbubble}).  The $s$-channel can be obtained easily from the $u$-channel result by continuing $p_-'\rightarrow -p_-'$  in $F^\mathrm{bubble}(p_-,p_-')$, but \emph{not} in $\Amp^{(0)}(p_-,p_-')$, resulting in
\[\label{sbubble}
\Amp^\mathrm{bubble}_s(\ppp) = \g^2 \Amp^{(0)}(p_-,p_-') \, F^\mathrm{bubble}(p_-,-p_-') \; .
\]
Combining the double bubbles together and substituting the expression for $I_{00}(\ppp)$ in (\ref{oneloopbubble00}), results in
\[ \label{stububble}
\begin{split}
\Amp_{stu}^\mathrm{bubbles} =\
    \g^2 \Amp^{(0)}(p_-,p_-') \Biggsbrk{ &
    - \frac{p_-^2 p_-'^2 (p_-'+p_-)^2}{(p_-'-p_-)^2}
    - \frac{2i p_-^2 p_-'^2 (p_-'+p_-)^2}{\pi(p_-'-p_-)^2} \ln\frac{p_-'}{p_-} \\ &
    + \frac{2  p_-^2 p_-'^2 (p_-^4+6p_-^2p_-'^2+p_-'^4)}{\pi^2 (p_-'^2-p_-^2)^2} \ln^2\frac{p_-'}{p_-}
    } \; .
\end{split}
\]

The wineglass diagrams are computationally more challenging because their loop integrals  do not factorize into products of one-loop integrals.  Nevertheless we are able to obtain compact expressions for these as well. Like the double bubble diagrams, all $2\rightarrow2$ processes have the same proportionality factor to their tree level amplitude.  For the $u$-channel wineglass, we find the expression
\[\label{uwine}
  \Amp_u^\mathrm{wine}(p,p') = 16 \g^2 \Amp^{(0)}(p_-,p_-') \, F^\mathrm{wine}(p_-,p_-') \; ,
\]
where
\[\label{Fwine}
\begin{split}
 F^\mathrm{wine}(\ppp) = 16 \Bigsbrk{
& - 4 p_-^2 p_-'^2 \, W_0(\ppp) + 8 p_- p_-' (p_-' + p_-) \, W_1(\ppp) \\
& - (p_-^2 +6 p_- p_-' + p_-'^2) \, W_2(\ppp) } \; .
\end{split}
\]
The wineglass integrals $W_r(\ppp)$ are defined and discussed in \appref{sec:wineglass}. Different processes have very different combinations to reach this same final form in (\ref{uwine}) and (\ref{Fwine}). The $s$-channel wineglass is again related to the $u$-channel form by analytically continuing $p'\to -p'$ in $F^\mathrm{wine}(\ppp) $,
\[\label{swine}
\Amp_s^\mathrm{wine}(p,p') = \g^2 \Amp^{(0)}(p_-,p_-') \, F^\mathrm{wine}(p_-,-p_-') \; .
\]
Likewise, we also find that the $t$-channel wineglass has a simple relation to the other wineglass diagrams, namely we simply set $p_-'=p_-$ in $F^\mathrm{wine}(\ppp)$, giving us
\[\label{twine}
\Amp_t^\mathrm{wine}(p,p') = \g^2 \Amp^{(0)}(p_-,p_-') \, F^\mathrm{wine}(p_-,p_-) \; .
\]

For the inverse wineglass diagrams, it is straightforward to show by the symmetries in the diagrams that
\be
\label{usinverse}
 \Amp_u^\mathrm{inverse}(\ppp) \eq \Amp_u^\mathrm{wine}(\ppp) \; , \nn \\
 \Amp_s^\mathrm{inverse}(\ppp) \eq \Amp_s^\mathrm{wine}(\ppp) \; ,
\ee
while the $t$-channel inverse wineglass is
\[\label{tinverse}
\Amp_t^\mathrm{inverse}(p,p') = \g^2 \Amp^{(0)}(p_-,p_-') \, F^\mathrm{wine}(p_-',p_-') \; .
\]

Putting together the terms in (\ref{uwine}), (\ref{swine}) and (\ref{usinverse}) and also using (\ref{Fwine}) and the expressions for $W_r(\ppp)$ in \appref{sec:wineglass} , we obtain  the combined $su$ wineglass
\[\label{uswine}
\begin{split}
\Amp_{su}^\mathrm{wineglasses} =\
   \g^2 \Amp^{(0)}(p_-,p_-') \Biggsbrk{ &
  - p_-^2 p_-'^2
  + \frac{8i\, p_-^3 p_-'^3}{\pi (p_-'^2-p_-^2)}
  + \frac{2i p_-^2 p_-'^2 (p_-^4+6p_-^2p_-'^2+p_-'^4)}{\pi   (p_-'^2-p_-^2)^2} \ln\frac{p_-'}{p_-} \\ &
  - \frac{2  p_-^2 p_-'^2 (p_-^4+6p_-^2p_-'^2+p_-'^4)}{\pi^2 (p_-'^2-p_-^2)^2} \ln^2\frac{p_-'}{p_-}
} \; .
\end{split}
\]
Combining the $t$-channel wineglass with its inverse gives
\[
\Amp_t^\mathrm{wineglasses} =\
  \g^2  \Amp^{(0)}(p_-,p_-') \biggsbrk{ \lrbrk{\frac{1}{\pi^2} - \frac{1}{12}} (p_-'^4 + p_-^4) } \; ,
\]
and then combining this with (\ref{stububble}) and (\ref{uswine}), we reach the final two-loop amplitude
\be\label{2loopamp}
  \Amp^{(2)}(p_-,p_-') \eq \Amp_{stu}^\mathrm{bubbles}(\ppp)
                          +\Amp_{su}^\mathrm{wineglasses}(\ppp)
                          +\Amp_t^\mathrm{wineglasses}(\ppp)\nn\\
  \eq \g^2\Amp^{(0)}(p_-,p_-')
  \Biggsbrk{
  - \frac{p_-^2 p_-'^2 (p_-'+p_-)^2}{(p_-'-p_-)^2}
  + \frac{8ip_-^3 p_-'^3}{\pi(p_-'^2 - p_-^2)}  \biggbrk{1 - \frac{p_-'^2 + p_-^2}{p_-'^2-p_-^2} \ln\frac{p_-'}{p_-} } \nn\\
&&\qquad\qquad\qquad\qquad - p_-^2 p_-'^2
  + \biggbrk{\frac{1}{\pi^2} - \frac{1}{12}}  (p_-'^4 + p_-^4)
} \; .
\ee

One should immediately note that the $(\ln \frac{p_-'}{p_-} )^2$ terms that appear in (\ref{stububble}) and (\ref{uswine}), but which are absent in the two-loop S-matrix in (\ref{limmatel},\ref{smatttr}) have canceled off in the final amplitude! One can also easily see that the first line of (\ref{2loopamp}) has precisely the right form as (\ref{smatttr2}). The first term in the second line is accounted for by a Jacobian factor, while the second term in this line, which is due entirely to the $t$-channel contributions, is compensated by wave-function renormalization of the external legs. In fact, the renormalization of the legs with momentum $p$ through them cancels off with the $t$-wineglass, while the renormalization of the $p'$ legs cancels against the inverse $t$-wineglass.

The Jacobian arises because the amplitudes come with factors of \mbox{$\delta^2(P^\mu_\mathrm{out}-P^\mu_\mathrm{in})$,} while the S-matrix is written with factors of $\delta(p_--q_-)\delta(p_-'-q_-')$.  These are related by
\be
\delta^2(P^\mu_\mathrm{out}-P^\mu_\mathrm{out})=\Half\left(\frac{dp_+'}{dp_-'}-\frac{dp_+}{dp_-}\right)^{-1}\delta(p_--q_-)\delta(p_-'-q_-') \; .
\ee
Taking into account the two-loop dispersion relation in \eqref{dispersion}, we find for the Jacobian
\begin{equation} \label{jacobian}
\Half \left(\frac{dp_+'}{dp_-'}-\frac{dp_+}{dp_-}\right)^{-1}=
 \frac{2p_-^2 p_-'^2}{m^2(p_-'^2-p_-^2)} \biggsbrk{1 + \frac{\g^2}{m^4} p_-^2p_-'^2 } \; .
\end{equation}
The full S-matrix has the form
\[
\Smatrix = \unit + \Half \left(\frac{dp_+'}{dp_-'}-\frac{dp_+}{dp_-}\right)^{-1}Z(p_-)Z(p_-') \, \Amp \; .
\]
Thus, after setting $m=1$ and substituting in \eqref{jacobian} and \eqref{wfren}, the two-loop contribution to the S-matrix is
\[
\Smatrix^{(2)} = \frac{p_- p_-'}{2(p_-'^2-p_-^2)} \Biggsbrk{ \Amp^{(2)} + \g^2 \Amp^{(0)}\biggbrk{p_-^2p_-'^2 - \biggbrk{ \frac{1}{\pi^2} - \frac{1}{12} } (p_-'^4 + p_-^4) }}
\]
Using the result for $\Amp^{(2)}$ in \eqref{2loopamp}, we reach the final expression
\[
\Smatrix^{(2)} = \g^2 \Amp^{(0)} \frac{p_-^3 p_-'^3}{2(p_-'^2-p_-^2)} \Biggsbrk{
  - \lrbrk{\frac{p_-'+p_-}{p_-'-p_-}}^2
  + \frac{8i}{\pi} \, \frac{p_- p_-'}{p_-'^2 - p_-^2} \lrbrk{1 - \frac{p_-'^2 + p_-^2}{p_-'^2-p_-^2} \ln\frac{p_-'}{p_-} } } \; ,
\]
which agrees precisely with the conjectured form \eqref{smatttr2}, since $\Smatrix^{(0)} = \Amp^{(0)} \frac{p_- p_-'}{2(p_-'^2-p_-^2)}$.

\section{Conclusions and outlook}

\label{sec:conclusion}

The sigma-model describing the super-string  on $\AdS_5\times\Sphere^5$ is a rather
complicated  theory and calculating the complete quantum S-matrix remains a formidable problem.
Fortunately consideration of the near-flat space limit, as described in \cite{Maldacena:2006rv}, results
in significant simplifications which make loop calculations feasible. The reduced sigma-model has at
most quartic interactions and the right movers essentially decouple from the interacting left-movers.
Just as for the full string theory in the light-cone gauge  the reduced model is not Lorentz invariant however if
boosts are combined with a rescaling of  the loop parameter the action is indeed invariant. This can be seen
in the world-sheet S-matrix which  depends only on the difference of rapidities and an effective, momentum
dependent, coupling.  Furthermore the simplified theory possesses at least  $(0,2)$ worldsheet supersymmetry.

As an important step in the calculation of
the S-matrix we computed the two-loop two point function with the corresponding mass shift and
wavefunction renormalisation. This is an interesting result in its own right as we can explicitly see
the modification of the relativistic dispersion by the sine function at higher powers of the momenta.
In the gauge theory description the sine function arises naturally from the intrinsic discreteness of the
spin-chain and indeed from the point of view of soliton description \cite{Hofman:2006xt} the momentum
is a periodic variable as it corresponds to the angular separation of the string endpoints. This is
however the first case where the sine function has been seen to originate from quantum
corrections to excitations about a plane-wave vacuum. Additionally in calculating the full $\grSU(2|2)$
S-matrix we are able to check that the symmetries of the classical theory are realized at higher loop order.

Given the central role of the world-sheet S-matrix in recent
developments of our understanding of the AdS/CFT correspondence it
certainly interesting to extract as much information and intuition
from this reduced model as possible. The spectacular agreement of
our calculations with the appropriate limit of the conjectured exact
S-matrix of
\cite{Beisert:2006ib},\cite{Beisert:2006ez}
provides further strong
evidence in favor of its validity. It should be straightforward,
though perhaps technically challenging, to extend the loop
calculation to even higher orders which would provide yet further
confirmation of the complete S-matrix. However, given that the
theory is presumably integrable, it may be more profitable to try to find a
complete solution using more non-perturbative techniques perhaps
along the lines discussed in \cite{Klose:2006dd}. This would allow
one to answer an outstanding issue not addressed by the perturbative
calculation, that of the pole structure of the S-matrix. Although we
consider the near-flat space limit which interpolates between the
plane wave limit and the giant magnon regime we do not see the
double poles of the S-matrix corresponding to exchange of BPS
magnons \cite{Dorey:2007xn}; which would require a resummation of
the entire perturbative expansion.

\paragraph{Note added} While this paper was being prepared for
publication we received \cite{new} where the study of two-loop
quantum corrections to the energies of classical string solutions
was initiated.

\bigskip
\subsection*{Acknowledgments}
\bigskip
We would like to thanks to J.~Maldacena, R. Roiban and  I.~Swanson
for discussions. 
The work of K.Z. was supported in part by the
Swedish Research Council under contracts 621-2004-3178 and
621-2003-2742, by grant NSh-8065.2006.2 for the support of
scientific schools, and by RFBR grant 06-02-17383. The work of T.K.
and K.Z. was supported by the G\"oran Gustafsson Foundation.  The work of J.A.M. was supported in part by the Swedish Research Council under contract 2006-3373.  J.A.M. and T.K. thank the CTP at MIT for hospitality during the course of this work, and the STINT foundation.

\appendix

\section{S-matrix elements}

\subsection{Bosons}

We write the action of the T-matrix, which is defined as $\Smatrix = \unit + \Tmatrix$, onto all bosonic initial states. We omit fermions in the final states. Using an $\grSO(4)\otimes\grSO(4)$ notation, we define the matrix elements as follows:
\begin{align}
\Tmatrix \ket{Z_i Z'_j} = \ &
    \ket{Z_i Z'_j} I_{ZZ}
  + \ket{Z_j Z'_i} P_{ZZ}
  + \delta_{ij} \ket{Z_k Z'_k} T_{ZZ}
  + \delta_{ij} \ket{Y_{k'} Y'_{k'}} K_{ZZ} \; , \\
\Tmatrix \ket{Z_i Y'_{j'}} = \ &
    \ket{Z_i Y'_{j'}} I_{ZY}
  + \ket{Y_{j'} Z'_i} P_{ZY} \; , \\
\Tmatrix \ket{Y_{i'} Z'_j} = \ &
    \ket{Y_{i'} Z'_j} I_{YZ}
  + \ket{Z_j Y'_{i'}} P_{YZ} \; , \\
\Tmatrix \ket{Y_{i'} Y'_{j'}} = \ &
    \ket{Y_{i'} Y'_{j'}} I_{YY}
  + \ket{Y_{j'} Y'_{i'}} P_{YY}
  + \delta_{i'j'} \ket{Y_{k'} Y'_{k'}} T_{YY}
  + \delta_{i'j'} \ket{Z_k Z'_k} K_{YY}
\end{align}
The world-sheet computation yields
\begin{align*}
 I_{ZZ}^{(0)} & = -I_{YY}^{(0)} = - 2i\g \, \frac{p_- p_-' (p_-'^2 + p_-^2)}{p_-'^2 - p_-^2} \\
 P_{ZZ}^{(0)} & = -P_{YY}^{(0)} = - 4i\g \, \frac{p_-^2 p_-'^2}{p_-'^2 - p_-^2} \\
 T_{ZZ}^{(0)} & = -T_{YY}^{(0)} = + 4i\g \, \frac{p_-^2 p_-'^2}{p_-'^2 - p_-^2} \\
 K_{ZZ}^{(0)} & = -K_{YY}^{(0)} = 0 \\[5mm]
 I_{ZZ}^{(1)} & = +I_{YY}^{(1)}
                = - 2\g^2 \, \frac{p_-^2 p_-'^2 (p_-'^2+p_-^2)}{(p_-'-p_-)^2}
                  + \frac{8i\g^2}{\pi} \, \frac{p_-^3 p_-'^3}{p_-'^2 - p_-^2}
                    \lrbrk{ 1 - \frac{p_-'^2+p_-^2}{p_-'^2-p_-^2} \ln\frac{p_-'}{p_-} } \\
 P_{ZZ}^{(1)} & = +P_{YY}^{(1)} = - 4\g^2 \, \frac{p_-^3 p_-'^3}{(p_-'-p_-)^2} \\
 T_{ZZ}^{(1)} & = +T_{YY}^{(1)} = + 4\g^2 \, \frac{p_-^3 p_-'^3}{(p_-'+p_-)^2} \\
 K_{ZZ}^{(1)} & = +K_{YY}^{(1)} = - 4\g^2 \, \frac{p_-^3 p_-'^3}{(p_-'+p_-)^2}
\\[5mm]
 I_{ZZ}^{(2)} & = -I_{YY}^{(2)}
                = + 2i\g^3 \, \frac{p_-^3 p_-'^3 (p_-'+p_-)(p_-'^2+p_-^2)}{(p_-'-p_-)^3}
                  + \frac{16\g^3}{\pi} \, \frac{p_-^4 p_-'^4 (p_-'^2+p_-^2)}{(p_-'^2-p_-^2)^2}
                    \lrbrk{ 1 - \frac{p_-'^2+p_-^2}{p_-'^2-p_-^2} \ln\frac{p_-'}{p_-} } \\
 P_{ZZ}^{(2)} & = -P_{YY}^{(2)}
                = + 4i\g^3 \, \frac{p_-^4 p_-'^4 (p_-'+p_-)}{(p_-'-p_-)^3}
                  + \frac{32\g^3}{\pi} \, \frac{p_-^5 p_-'^5}{(p_-'^2-p_-^2)^2}
                    \lrbrk{ 1 - \frac{p_-'^2+p_-^2}{p_-'^2-p_-^2} \ln\frac{p_-'}{p_-} } \\
 T_{ZZ}^{(2)} & = -T_{YY}^{(2)}
                = - 4i\g^3 \, \frac{p_-^4 p_-'^4 (p_-'+p_-)}{(p_-'-p_-)^3}
                  - \frac{32\g^3}{\pi} \, \frac{p_-^5 p_-'^5}{(p_-'^2-p_-^2)^2}
                    \lrbrk{ 1 - \frac{p_-'^2+p_-^2}{p_-'^2-p_-^2} \ln\frac{p_-'}{p_-} } \\
 K_{ZZ}^{(2)} & = -K_{YY}^{(2)} = 0
\end{align*}
\begin{align*}
 I_{ZY}^{(0)} & = - I_{YZ}^{(0)} = - 2i\g \, p_- p_-' \\
 P_{ZY}^{(0)} & = - P_{ZY}^{(0)} = 0 \\[5mm]
 I_{ZY}^{(1)} & = + I_{YZ}^{(1)}
                = - 2\g^2 \, \frac{p_-^2 p_-'^2 (p_-'^2+p_-^2)}{(p_-'-p_-)^2}
                  + \frac{8i\g^2}{\pi} \frac{p_-^3 p_-'^3}{p_-'^2-p_-^2}
                    \lrbrk{ 1 - \frac{p_-'^2+p_-^2}{p_-'^2-p_-^2} \ln\frac{p_-'}{p_-} } \\
 P_{ZY}^{(1)} & = + P_{YZ}^{(1)}
                = - 4\g^2 \, \frac{p_-^3 p_-'^3}{(p_-'-p_-)^2} \\[5mm]
 I_{ZY}^{(2)} & = - I_{YZ}^{(2)}
                = + 2i\g^3 \, \frac{p_-^3 p_-'^3 (p_-'+p_-)^2}{(p_-'-p_-)^2}
                  + \frac{16\g^3}{\pi} \, \frac{p_-^4 p_-'^4}{p_-'^2-p_-^2}
                    \lrbrk{ 1 - \frac{p_-'^2+p_-^2}{p_-'^2-p_-^2} \ln\frac{p_-'}{p_-} } \\
 P_{ZY}^{(2)} & = - P_{YZ}^{(2)} = 0
\end{align*}
These coefficients have to be compared to the S-matrix elements \eqref{smatttr} in the follow way:
\begin{align}
1+I_{YY} & =   S_0 \, A \, (A+B) \; , &
1+I_{ZZ} & =   S_0 \, D \, (D+E) \; , \\
  P_{YY} & =   S_0 \, B \, (A+B) \; , &
  P_{ZZ} & =   S_0 \, E \, (D+E) \; , \nn \\
  T_{YY} & = - S_0 \, A \, B     \; , &
  T_{ZZ} & = - S_0 \, D \, E     \; , \nn \\
  K_{YY} & = - S_0 \, C^2        \; , &
  K_{ZZ} & = - S_0 \, F^2        \; , \nn \\[2mm]
1+I_{ZY} & =   S_0 \, L^2        \; , &
1+I_{YZ} & =   S_0 \, G^2        \; , \nn \\
  P_{ZY} & =   S_0 \, K^2        \; , &
  P_{YZ} & =   S_0 \, H^2        \; . \nn
\end{align}
Here $S_0$ denotes the prefactor in \eqref{Smtr}. We find perfect agreement. Note that we are sensitive to \emph{all} matrix elements, even though we concentrate onto the scattering among bosons. This is because the field $Z$ actually carries two fermionic indices in the $\algSU(2|2)^2$ notation.

\subsection{$\grSU(2|2)$ subsector} \label{sec:su22}

We now extend our considerations to include processes involving fermions, however, we restrict ourselves to a single $\algSU(2|2)$ sector. Granting the group factorization of the full S-matrix, this is a sufficient test of the supersymmetries at higher loop orders.

As described in \appref{sec:indicies}, we identify in the worldsheet theory the fields $\phi_a$ and $\chi_\alpha$ spanning an $\algSU(2|2)$ sector. We calculate the matrix elements defined as follows:
\begin{align}
\Smatrix\ket{\phi_a\phi'_b} &=
{\cal S}_0 {\cal \Asmatrix}\ket{\phi_{a}\phi'_{b}}
+{\cal S}_0{\cal \Bsmatrix}\ket{\phi_{b}\phi'_{a}}
+{\cal S}_0 {\cal \Csmatrix} \varepsilon_{ab}\varepsilon^{\alpha\beta}\ket{\chi_\alpha\chi'_\beta} \; ,
\\
\Smatrix \ket{\chi_\alpha\chi'_\beta} &=
{\cal S}_0{\cal \Dsmatrix} \ket{\chi_{\alpha}\chi'_{\beta}}
+{\cal S}_0{\cal \Esmatrix} \ket{\chi_{\beta}\chi'_{\alpha}}
+{\cal S}_0 {\cal \Fsmatrix }\varepsilon_{\alpha\beta}\varepsilon^{ab}\ket{\phi_a\phi'_b} \; ,
\\
\Smatrix\ket{\phi_a\chi'_\beta} &=
{\cal S}_0{\cal  \Gsmatrix}\ket{\phi_a\chi'_{\beta}}
+{\cal S}_0{\cal \Hsmatrix}\ket{\chi_\beta \phi'_{a}} \; ,
\\
\Smatrix\ket{\chi_\alpha\phi'_b} &=
{\cal S}_0 {\cal \Ksmatrix}\ket{\phi_{b}\chi'_\alpha}
+{\cal S}_0{\cal \Lsmatrix}\ket{\chi_\alpha\phi'_{b}} \; .
\label{SU22_S_matrix_simp}
\end{align}

For the sake of brevity we will not record the individual amplitudes but simply state the final results for the $\smatrix$-matrix elements, noting that they agree with the all-order prediction from the dual
spin chain description. There is a common
contribution to each element of the form ${\cal S}_0=1+i \delta$ where
\[
\delta = \frac{8 \g}{\pi} \frac{p_-^3p_-'^3}{p_-'^2-p_-^2} \left(1-\frac{p_-'^2+p_-^2}{p_-'^2-p_-^2} \ln \frac{p_-'}{p_-}\right)
\]
in addition to the individual contributions
\be
{\cal \Asmatrix}&=&
1
+2i\g \, \frac{p_- p_-'(p_-^2+p_-'^2)}{p_-'^2-p_-^2}
-2\g^2 \, \frac{p_-^2p_-'^2(p_-^2+p_-'^2)}{(p_-'-p_-)^2}
+2i\g^3 \, \frac{p_-^3p_-'^3(p_-^3+p_-^2p_-'+p_- p_-'^2+p_-'^3)}{(p_-'-p_-)^3}\nn\\
{\cal \Bsmatrix}&=&-\Esmatrix=
4i\g \, \frac{p_-^2 p_-'^2}{p_-'^2-p_-^2}
-4\g^2 \, \frac{p_-^3p_-'^3}{(p_-'-p_-)^2}
-4i\g^3 \, \frac{p_-^4p_-'^4(p_-+p_-')}{(p_-'-p_-)^3}\nn\\
{\cal \Csmatrix}&=&{\cal \Fsmatrix}=
2i\g \, \frac{(p_- p_-')^{\frac{3}{2}}}{p_-'+p_-}
-2\g^2 \, \frac{(p_-p_-')^{\frac{5}{2}}}{(p_-'-p_-)}
-2i\g^3 \, \frac{(p_-p_-')^{\frac{7}{2}}(p_-+p_-')}{(p_-'-p_-)^2}\nn\\
{\cal \Dsmatrix}&=&
1
+4i\g \, \frac{p_-^2 p_-'^2}{p_-'^2-p_-^2}
-4\g^2 \, \frac{p_-^3p_-'^3}{(p_-'-p_-)^2}
-4i\g^3 \, \frac{p_-^4p_-'^4(p_-+p_-')}{(p_-'-p_-)^3}\nn\\
{\cal \Gsmatrix}&=&
1
+2i\g \, \frac{p_- p_-'^2}{p_-'-p_-}
-2\g^2 \, \frac{p_-^2p_-'^3(p_-+p_-')}{(p_-'-p_-)^2}
-2i\g^3 \, \frac{p_-^3p_-'^4(p_-+p_-)^2}{(p_-'-p_-)^3}\nn\\
{\cal \Hsmatrix}&=&{\cal \Ksmatrix}=
2i\g \, \frac{(p_- p_-')^\frac{3}{2}}{p_-'-p_-}
+2\g^2 \, \frac{(p_-p_-')^\frac{5}{2}(p_-+p_-')}{(p_-'-p_-)^2}
-2i\g^3 \, \frac{(p_-p_-')^\frac{7}{2}(p_-+p_-')^2}{(p_-'-p_-)^3}\nn\\
{\cal \Lsmatrix}&=&
1
+2i\g \, \frac{p_-^2 p_-'}{p_-'-p_-}
-2\g^2 \, \frac{p_-^3p_-'^2(p_-+p_-')}{(p_-'-p_-)^2}
-2i\g^3 \, \frac{p_-^4p_-'^3(p_-+p_-)^2}{(p_-'-p_-)^3}\ .\nn
\ee
In considering a single $\grSU(2|2)$ sector the full S-matrix is
\[
\Smatrix = S_0\,\smatrix\otimes (\Asmatrix+\Bsmatrix)
\]
as one index in the tensor product is kept fixed by the scattering.
Thus we can write these elements in a simple compact form in terms of the S-matrix defined in \ref{limmatel},
\[
\Smatrix= \frac{e^{i\delta} }{1-i\g p_- p_-'\frac{p_-'+p_-}{p_-'-p_-}} \ \smatrix\otimes \unit \; ,
\]
and which of course is in agreement with the AdS/CFT prediction to this order. Thus we see that the
symmetries are preserved to at least two-loops in the reduced sigma model.

\section{Integrals}

\begin{figure}
\begin{center}
\subfigure[bubble]{\includegraphics[scale=0.8]{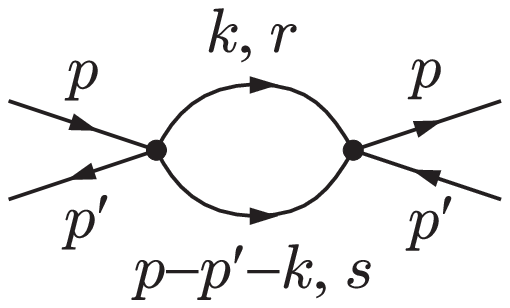}\label{fig:bubble}}\qquad
\subfigure[sunset]{\includegraphics[scale=0.8]{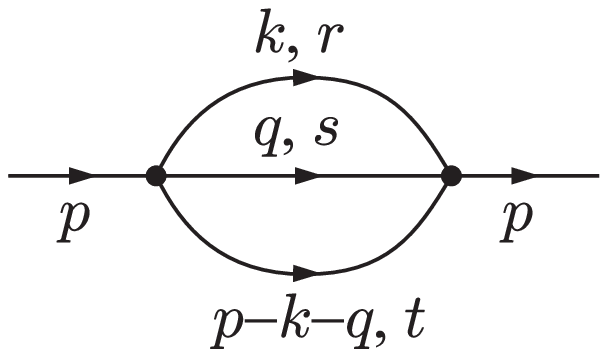}\label{fig:sunset}}\qquad
\subfigure[wineglass]{\includegraphics[scale=0.8]{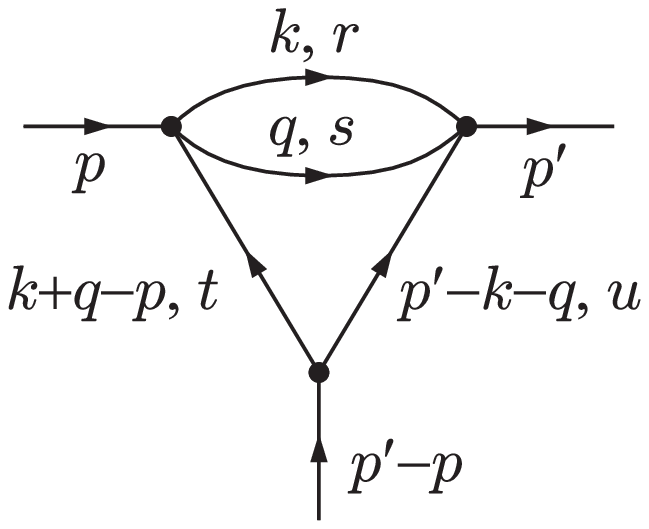}\label{fig:wineglass}}
\end{center}
\caption{\textbf{Loop diagrams}}
\end{figure}

\subsection{Bubble integral}
\label{app:bubble}

We consider the bubble integral, cf. \figref{fig:bubble}, for two inflowing momenta $p$ and $-p'$ as appropriate for $u$-channel processes. With $r$ and $s$ powers of momentum inserted, the integral reads
\[ \label{oneloopbubble}
  I_{rs}(p,p') = \int \frac{d^2\vec{k}}{(2\pi)^2} \frac{(k_-)^r (p_--p_-'-k_-)^s}{(\vec{k}^2 - m^2)[(\vec{p}-\vec{p}'-\vec{k})^2 - m^2]} \; .
\]
These momenta originate from derivative couplings and fermionic propagators. However, it turns out that all amplitudes simplify such that we only need to explicitely compute $I_{00}$, which is immediately found to be
\[ \label{oneloopbubble00}
I_{00}(p,p') = \frac{i}{2\pi m^2} \frac{p_- p_-'}{p_-'^2 - p_-^2} \ln\frac{p_-'}{p_-} \; .
\]
In the $s$-channel, the inflowing momenta are $p$ and $p'$. The integral $I_{00}(p,-p')$ is obtained from \eqref{oneloopbubble00} by analytically continuing the logarithm. In the $t$-channel, the total inflowing momentum is zero and we obtain from \eqref{oneloopbubble00} in the limit $p'\to p$:
\[
  I_{00}(p,p) = \frac{i}{4\pi m^2} \; .
\]

\subsection{Sunset integral}
\label{app:sunset}

The general sunset diagram, \figref{fig:sunset}, is defined as
\[
I_{rst}(p) = \int \frac{d^2\vec{k}\,d^2\vec{q}}{(2\pi)^4} \frac{(k_-)^r \, (q_-)^s \, (p_- - k_- - q_-)^t}{(\vec{k}^2 - m^2)(\vec{q}^2 - m^2)((\vec{p}-\vec{k}-\vec{q})^2 - m^2)} \; .
\]
There is the relation
\[ \label{eqn:Irst-identity}
  p_- I_{rst} = I_{r+1,s,t} + I_{r,s+1,t} + I_{r,s,t+1}
\]
between different integrals which follows immediately from taking the $p_-$ on the left hand side into the integrand and writing it as $k_- + q_- + (p_- - k_- - q_-)$. Using this identity, it is possible to reduce all sums of sunset diagrams that occur in the two-loop propagator to $I_{000}$. We solve this integral by introducing three Feynman parameters
\[
I_{000}(\vec{p}^2) = \frac{1}{16\pi^2} \int_0^1
\frac{dx_1\, dx_2 \, dx_3 \; \delta(x_1+x_2+x_3-1)}{m^2 (x_1 x_2 + x_1 x_3 + x_2 x_3) - \vec{p}^2 \, x_1 x_2 x_3} \; .
\]
Observe that this integral depends only on $\vec{p}^2$. On-shell the value of the integral is
\[ \label{eqn:I000-on-shell}
I_{000}(\vec{p}^2=m^2) = \frac{1}{64m^2} \; .
\]
Apart form this, we also need the on-shell value of the first derivative of $I_{000}$ with repect to its argument, which is given by
\[ \label{eqn:I000prime}
\begin{split}
I'_{000}(\vec{p}^2=m^2) & = \frac{1}{(4\pi)^2 m^4} \int_0^1
\frac{dx_1\, dx_2 \, dx_3 \; \delta(x_1+x_2+x_3-1) \, x_1 x_2 x_3}{ (x_1 x_2 + x_1 x_3 + x_2 x_3 - x_1 x_2 x_3)^2 } \\
& = \frac{3}{64m^4} \lrbrk{\frac{1}{\pi^2}-\frac{1}{12}} \; .
\end{split}
\]

\subsection{Wineglass integral}
\label{sec:wineglass}

The wineglass diagram as drawn in \figref{fig:wineglass} reads
\[
W_{rstu}(p,p') = \int\frac{d^2\vec{k}\,d^2\vec{q}}{(2\pi)^4}\, \frac{(k_-)^{n_1}(q_-)^{n_2}(k_-+q_--p_-)^{n_3}(p'_--k_--q_-)^{n_4}}{(\vec{k}^2-m^2)(\vec{q}^2-m^2)[(\vec{k}+\vec{q}-\vec{p})^2-m^2][(\vec{p}'-\vec{k}-\vec{q})^2-m^2]} \; .
\]
We note the identities
\[
 W_{rstu}(p,p') = W_{srtu}(p,p') = (-1)^{t+u} W_{rsut}(p',p) = (-1)^{r+s+t+u} W_{rstu}(-p,-p') \; .
\]
All sums of wineglass integrals that occur in the two-loop amplitudes can be reduced to combinations of the following three terms which we compute by standard means and find
\begin{align}
W_0(p,p') & = W_{0000}(p,p') \nn \\
          & = - \frac{p_- p_-'}{16\pi^2m^4} \lrsbrk{ \frac{\pi^2}{4(p_-'+p_-)^2}
              + \frac{1}{(p_-'^2 - p_-^2)} \ln\frac{p_-'}{p_-}
              - \frac{p_- p_-'}{(p_-'^2 - p_-^2)^2} \ln^2\frac{p_-'}{p_-} } \; , \\
W_1(p,p') & = W_{1000}(p,p') \nn \\
          & = - \frac{p_- p_-'}{16\pi^2m^4} \lrsbrk{ \frac{\pi^2}{8(p_-'+p_-)}
              - \frac{p_- p_-'}{2(p_-' - p_-)(p_-'^2 - p_-^2)} \ln^2\frac{p_-'}{p_-} } \; , \\
W_2(p,p') & = W_{1100}(p,p') + W_{2000}(p,p') \nn \\
          & = - \frac{p_- p_-'}{16\pi^2m^4} \lrsbrk{ \frac{\pi^2}{12}
              - \frac{p_- p_-'}{2 (p_-' - p_-)^2} \ln^2\frac{p_-'}{p_-} } \; .
\end{align}

\section{Notations}
\label{sec:indicies}

In this section we summarize several of the notations used throughout
the main text and record several useful results.
We make use of the light-cone coordinates and momenta
\[ \label{eqn:lc-momenta}
\sigma^\pm = \sigma^0 \pm \sigma^1
\comma
p_\pm = \half (p_0 \pm p_1) \;
\]
so that the worldsheet metric is  $\eta_{\mu\nu} = (+,-)$.
We also use the notation $\energy = p_0$ and $p = p_1$, and bold-face for world-sheet two-vectors like $\vec{p} = (p_0,p_1)$.

It is convenient to perform quantization in world-sheet light-cone coordinates
with $\sigma^+$ as time and where the  target space fields have the mode expansions
\begin{align}
Z_i(\vecsigma)    & = \int\frac{dp_-}{2\pi} \frac{1}{\sqrt{2p_-}} \:
                    \Bigsbrk{ a_i(p_-)      \, e^{-i\vec{p}\cdot\vecsigma}
                            + a^\dag_i(p_-) \, e^{+i\vec{p}\cdot\vecsigma} } \; , \\
Y_{i'}(\vecsigma) & = \int\frac{dp_-}{2\pi} \frac{1}{\sqrt{2p_-}} \:
                 \Bigsbrk{ a_{i'}(p_-)      \, e^{-i\vec{p}\cdot\vecsigma}
                         + a^\dag_{i'}(p_-) \, e^{+i\vec{p}\cdot\vecsigma} } \; , \\
\psi(\vecsigma) & = \int\frac{dp_-}{2\pi} \frac{1}{\sqrt{2}} \:
                  \Bigsbrk{ b(p_-)      \, e^{-i\vec{p}\cdot\vecsigma}
                          + b^\dag(p_-) \, e^{+i\vec{p}\cdot\vecsigma} } \; .
\end{align}
\noindent
The free bosonic and fermionic propagators are
\[
\frac{i}{\vec{p}^2 - m^2} \comma \frac{ip_-}{\vec{p}^2 - m^2}
\]
and the free dispersion relation is $2 p_+ =\tfrac {m^2}{2 p_-} $.

We use the following representation for the $16\times 16$ $\gamma$-matrices
\begin{align}
\Gamma^1 & = \epsilon \times \epsilon\times \epsilon\times \epsilon &
\Gamma^5 & = \tau_3   \times \epsilon\times \unit   \times \epsilon \nn\\
\Gamma^2 & = \unit    \times \tau_1  \times \epsilon\times \epsilon &
\Gamma^6 & = \epsilon \times \unit   \times \tau_1  \times \epsilon \label{cliffmat} \\
\Gamma^3 & = \unit    \times \tau_3  \times \epsilon\times \epsilon &
\Gamma^7 & = \epsilon \times \unit   \times \tau_3  \times \epsilon \nn\\
\Gamma^4 & = \tau_1   \times \epsilon\times \unit   \times \epsilon &
\Gamma^8 & = \unit    \times \unit   \times \unit   \times \tau_1   \nn
\end{align}
with
\be
\epsilon = \matr{cc}{ 0 & 1 \\ -1 & 0 } \comma
\tau_1   = \matr{cc}{ 0 & 1 \\  1 & 0 } \comma
\tau_3   = \matr{cc}{ 1 & 0 \\ 0 & -1 } \; .
\ee
We also define $\Gamma^9 = \Gamma^1 \Gamma^2 \cdots \Gamma^8$ and $P_{L,R} = \half(\unit\pm\Gamma^9)$. The fermion $\psi$ is a real, positive chirality spinor and hence has eight real degrees of freedom.

The $\algSU(2|2)$ sector considered in \secref{sec:su22} is spanned by the bosonic fields
\begin{align}
  \phi_1 & = \frac{1}{\sqrt{2}} (Y_5 + i Y_6) \; , &
  \phi_2 & = \frac{1}{\sqrt{2}} (Y_7 + i Y_8) \; ,
\end{align}
and the fermionic fields, $\chi_\alpha$, which are most easily defined in terms of the projection operators,
\be
P^{I}_\pm&=&\frac{1}{2}\left(\unit\pm\Gamma^1\Gamma^2\Gamma^3\Gamma^4\right)\nn\\
P^{II}_\pm&=&\frac{1}{2}\left(\unit\pm\frac{i}{2}\left(\Gamma^{56}+\Gamma^{78}\right)\right)\nn\\
P^{III}_\pm&=&\frac{1}{2}\left(\unit\pm\frac{i}{2}\left(\Gamma^{12}+\Gamma^{34}\right)\right),
\ee
such that,
\be
 \chi_3 & = &P^{III}_+P^{II}_+P^{I}_-P_{L}\psi\; , \nn\\
  \chi_4 & =  &P^{III}_-P^{II}_+P^{I}_-P_{L}\psi\ \; .
\ee
These correspond to the fields $\phi_a = Y_{a\dot{1}}$ and $\chi_\alpha = \Bsi_{\alpha\dot{1}}$ in the notation of \cite{Klose:2006zd}.

\section{S-matrix action}
\label{app:everything}

We spell out the action of the S-matrix onto the entire set of two-particle states in $\algSU(2)^4$ notation, cf.~\cite{Klose:2007wq}. This serves as a reference for which processes can occur. Taking into account that the coefficients $B$, $E$, $C$, $F$, $H$, $K$ are of order $\g = \frac{\pi}{\sqrt{\lambda}}$, we see that some of the processes are absent at tree-level. The terms that are present at tree-level are printed in bold face. To simplify the formulas, we suppress the $S_0$ that multiplies all right hand sides in the following.

\newcommand{\treelevel}[1]{\mbox{\mathversion{bold}$#1$}}

\subsubsection*{Boson-Boson}
\[ \nn
\begin{split}
\Smatrix \ket{Y_{\lAA \rAA} Y'_{\lBB \rBB}} = \ &
+ A^2 \, \treelevel{\ket{Y_{\lAA \rAA} Y'_{\lBB \rBB}}}
+ A B \, \treelevel{\ket{Y_{\lBB \rAA} Y'_{\lAA \rBB}}}
+ A B \, \treelevel{\ket{Y_{\lAA \rBB} Y'_{\lBB \rAA}}}
+ B^2 \, \ket{Y_{\lBB \rBB} Y'_{\lAA \rAA}} \\ &
+ A C \, \levi_{\lAA \lBB} \levi^{\lcc \ldd} \treelevel{\ket{\Bsi_{\lcc \rAA} \Bsi'_{\ldd \rBB}}}
+ B C \, \levi_{\lAA \lBB} \levi^{\lcc \ldd} \ket{\Bsi_{\lcc \rBB} \Bsi'_{\ldd \rAA}} \\ &
+ A C \, \levi_{\rAA \rBB} \levi^{\rcc \rdd} \treelevel{\ket{\Psi_{\lAA \rcc} \Psi'_{\lBB \rdd}}}
+ B C \, \levi_{\rAA \rBB} \levi^{\rcc \rdd} \ket{\Psi_{\lBB \rcc} \Psi'_{\lAA \rdd}} \\ &
+ C^2 \, \levi_{\lAA \lBB} \levi_{\rAA \rBB} \levi^{\lcc \ldd} \levi^{\rcc \rdd} \ket{Z_{\lcc \rcc} Z'_{\ldd \rdd}} \\[3mm]
\Smatrix \ket{Z_{\laa \raa} Z'_{\lbb \rbb}} = \ &
+ D^2 \, \treelevel{\ket{Z_{\laa \raa} Z'_{\lbb \rbb}}}
+ D E \, \treelevel{\ket{Z_{\lbb \raa} Z'_{\laa \rbb}}}
+ D E \, \treelevel{\ket{Z_{\laa \rbb} Z'_{\lbb \raa}}}
+ E^2 \, \ket{Z_{\lbb \rbb} Z'_{\laa \raa}} \\ &
+ D F \, \levi_{\raa \rbb} \levi^{\rCC \rDD} \treelevel{\ket{\Bsi_{\laa \rCC} \Bsi'_{\lbb \rDD}}}
+ E F \, \levi_{\raa \rbb} \levi^{\rCC \rDD} \ket{\Bsi_{\lbb \rCC} \Bsi'_{\laa \rDD}} \\ &
+ D F \, \levi_{\laa \lbb} \levi^{\lCC \lDD} \treelevel{\ket{\Psi_{\lCC \raa} \Psi'_{\lDD \rbb}}}
+ E F \, \levi_{\laa \lbb} \levi^{\lCC \lDD} \ket{\Psi_{\lCC \rbb} \Psi'_{\lDD \raa}} \\ &
+ F^2 \, \levi_{\laa \lbb} \levi_{\raa \rbb} \levi^{\lCC \lDD} \levi^{\rCC \rDD} \ket{Y_{\lCC \rCC} Y'_{\lDD \rDD}} \\[3mm]
\Smatrix \ket{Y_{\lAA \rAA} Z'_{\lbb \rbb}} = \ &
+ G^2 \, \treelevel{\ket{Y_{\lAA \rAA} Z'_{\lbb \rbb}}}
+ G H \, \treelevel{\ket{\Bsi_{\lbb \rAA} \Psi'_{\lAA \rbb}}}
+ G H \, \treelevel{\ket{\Psi_{\lAA \rbb} \Bsi'_{\lbb \rAA}}}
+ H^2 \, \ket{Z_{\lbb \rbb} Y'_{\lAA \rAA}} \\[3mm]
\Smatrix \ket{Z_{\laa \raa} Y'_{\lBB \rBB}} = \ &
+ L^2 \, \treelevel{\ket{Z_{\laa \raa} Y'_{\lBB \rBB}}}
+ K L \, \treelevel{\ket{\Psi_{\lBB \raa} \Bsi'_{\laa \rBB}}}
+ K L \, \treelevel{\ket{\Bsi_{\laa \rBB} \Psi'_{\lBB \raa}}}
+ K^2 \, \ket{Y_{\lBB \rBB} Z'_{\laa \raa}}
\end{split}
\]

\vspace{4mm}

\subsubsection*{Fermion-Fermion}
\[ \nn
\begin{split}
\Smatrix \ket{\Psi_{\lAA \raa} \Psi'_{\lBB \rbb}} = \ &
+ A D \, \treelevel{\ket{\Psi_{\lAA \raa} \Psi'_{\lBB \rbb}}}
+ B D \, \treelevel{\ket{\Psi_{\lBB \raa} \Psi'_{\lAA \rbb}}}
+ A E \, \treelevel{\ket{\Psi_{\lAA \rbb} \Psi'_{\lBB \raa}}}
+ B E \, \ket{\Psi_{\lBB \rbb} \Psi'_{\lAA \raa}} \\ &
+ A F \, \levi_{\rCC \rDD} \levi^{\raa \rbb} \treelevel{\ket{Y_{\lAA \rCC} Y'_{\lBB \rDD}}}
+ B F \, \levi_{\rCC \rDD} \levi^{\raa \rbb} \ket{Y_{\lBB \rCC} Y'_{\lAA \rDD}} \\ &
+ C D \, \levi_{\lAA \lBB} \levi^{\lcc \ldd} \treelevel{\ket{Z_{\lcc \raa} Z'_{\ldd \rbb}}}
+ C E \, \levi_{\lAA \lBB} \levi^{\lcc \ldd} \ket{Z_{\lcc \rbb} Z'_{\ldd \raa}} \\ &
+ C F \, \levi_{\lAA \lBB} \levi_{\raa \rbb} \levi^{\lcc \ldd} \levi^{\rCC \rDD} \ket{\Bsi_{\lcc \rCC} \Bsi'_{\ldd \rDD}} \\[3mm]
\Smatrix \ket{\Bsi_{\laa \rAA} \Bsi'_{\lbb \rBB}} = \ &
+ A D \, \treelevel{\ket{\Bsi_{\laa \rAA} \Bsi'_{\lbb \rBB}}}
+ A E \, \treelevel{\ket{\Bsi_{\lbb \rAA} \Bsi'_{\laa \rBB}}}
+ B D \, \treelevel{\ket{\Bsi_{\laa \rBB} \Bsi'_{\lbb \rAA}}}
+ B E \, \ket{\Bsi_{\lbb \rBB} \Bsi'_{\laa \rAA}} \\ &
+ A F \, \levi_{\laa \lbb} \levi^{\lCC \lDD} \treelevel{\ket{Y_{\lCC \rAA} Y'_{\lDD \rBB}}}
+ B F \, \levi_{\laa \lbb} \levi^{\lCC \lDD} \ket{Y_{\lCC \rBB} Y'_{\lDD \rAA}} \\ &
+ C D \, \levi_{\rAA \rBB} \levi^{\rcc \rdd} \treelevel{\ket{Z_{\laa \rcc} Z'_{\lbb \rdd}}}
+ C E \, \levi_{\rAA \rBB} \levi^{\rcc \rdd} \ket{Z_{\lbb \rcc} Z'_{\laa \rdd}} \\ &
+ C F \, \levi_{\rAA \rBB} \levi_{\laa \lbb} \levi^{\lCC \lDD} \levi^{\rcc \rdd} \ket{\Psi_{\lCC \rcc} \Psi'_{\lDD \rdd}} \\[3mm]
\Smatrix \ket{\Psi_{\lAA \raa} \Bsi'_{\lbb \rBB}} = \ &
+ G L \, \treelevel{\ket{\Psi_{\lAA \raa} \Bsi'_{\lbb \rBB}}}
+ H L \, \treelevel{\ket{Z_{\lbb \raa} Y'_{\lAA \rBB}}}
+ G K \, \treelevel{\ket{Y_{\lAA \rBB} Z'_{\lbb \raa}}}
+ H K \, \ket{\Bsi_{\lbb \rBB} \Psi'_{\lAA \raa}} \\[3mm]
\Smatrix \ket{\Bsi_{\laa \rAA} \Psi'_{\lBB \rbb}} = \ &
+ G L \, \treelevel{\ket{\Bsi_{\laa \rAA} \Psi'_{\lBB \rbb}}}
+ G K \, \treelevel{\ket{Y_{\lBB \rAA} Z'_{\laa \rbb}}}
+ H L \, \treelevel{\ket{Z_{\laa \rbb} Y'_{\lBB \rAA}}}
+ H K \, \ket{\Psi_{\lBB \rbb} \Bsi'_{\laa \rAA}}
\end{split}
\]

\subsubsection*{Boson-Fermion}
\[ \nn
\begin{split}
\Smatrix \ket{Y_{\lAA \rAA} \Psi'_{\lBB \rbb}} = \ &
+ A G \, \treelevel{\ket{Y_{\lAA \rAA} \Psi'_{\lBB \rbb}}}
+ B G \, \treelevel{\ket{Y_{\lBB \rAA} \Psi'_{\lAA \rbb}}}
+ A H \, \treelevel{\ket{\Psi_{\lAA \rbb} Y'_{\lBB \rAA}}}
+ B H \, \ket{\Psi_{\lBB \rbb} Y'_{\lAA \rAA}} \\ &
+ C G \, \levi_{\lAA \lBB} \levi^{\lcc \ldd} \treelevel{\ket{\Bsi_{\lcc \rAA} Z'_{\ldd \rbb}}}
+ C H \, \levi_{\lAA \lBB} \levi^{\lcc \ldd} \ket{Z_{\lcc \rbb} \Bsi'_{\ldd \rAA}} \\[3mm]
\Smatrix \ket{Y_{\lAA \rAA} \Bsi'_{\lbb \rBB}} = \ &
+ A G \, \treelevel{\ket{Y_{\lAA \rAA} \Bsi'_{\lbb \rBB}}}
+ A H \, \treelevel{\ket{\Bsi_{\lbb \rAA} Y'_{\lAA \rBB}}}
+ B G \, \treelevel{\ket{Y_{\lAA \rBB} \Bsi'_{\lbb \rAA}}}
+ B H \, \ket{\Bsi_{\lbb \rBB} Y'_{\lAA \rAA}} \\ &
+ C G \, \levi_{\rAA \rBB} \levi^{\rcc \rdd} \treelevel{\ket{\Psi_{\lAA \rcc} Z'_{\lbb \rdd}}}
+ C H \, \levi_{\rAA \rBB} \levi^{\rcc \rdd} \ket{Z_{\lbb \rcc} \Psi'_{\lAA \rdd}} \\[3mm]
\Smatrix \ket{\Psi_{\lAA \raa} Y'_{\lBB \rBB}} = \ &
+ A L \, \treelevel{\ket{\Psi_{\lAA \raa} Y'_{\lBB \rBB}}}
+ B L \, \treelevel{\ket{\Psi_{\lBB \raa} Y'_{\lAA \rBB}}}
+ A K \, \treelevel{\ket{Y_{\lAA \rBB} \Psi'_{\lBB \raa}}}
+ B K \, \ket{Y_{\lBB \rBB} \Psi'_{\lAA \raa}} \\ &
+ C L \, \levi_{\lAA \lBB} \levi^{\lcc \ldd} \treelevel{\ket{Z_{\lcc \raa} \Bsi'_{\ldd \rBB}}}
+ C K \, \levi_{\lAA \lBB} \levi^{\lcc \ldd} \ket{\Bsi_{\lcc \rBB} Z'_{\ldd \raa}} \\[3mm]
\Smatrix \ket{\Bsi_{\laa \rAA} Y'_{\lBB \rBB}} = \ &
+ A L \, \treelevel{\ket{\Bsi_{\laa \rAA} Y'_{\lBB \rBB}}}
+ A K \, \treelevel{\ket{Y_{\lBB \rAA} \Bsi'_{\laa \rBB}}}
+ B L \, \treelevel{\ket{\Bsi_{\laa \rBB} Y'_{\lBB \rAA}}}
+ B K \, \ket{Y_{\lBB \rBB} \Bsi'_{\laa \rAA}} \\ &
+ C L \, \levi_{\rAA \rBB} \levi^{\rcc \rdd} \treelevel{\ket{Z_{\laa \rcc} \Psi'_{\lBB \rdd}}}
+ C K \, \levi_{\rAA \rBB} \levi^{\rcc \rdd} \ket{\Psi_{\lBB \rcc} Z'_{\laa \rdd}}
\end{split}
\]

\vspace{2mm}

\[ \nn
\begin{split}
\Smatrix \ket{Z_{\laa \raa} \Psi'_{\lBB \rbb}} = \ &
+ D L \, \treelevel{\ket{Z_{\laa \raa} \Psi'_{\lBB \rbb}}}
+ D K \, \treelevel{\ket{\Psi_{\lBB \raa} Z'_{\laa \rbb}}}
+ E L \, \treelevel{\ket{Z_{\laa \rbb} \Psi'_{\lBB \raa}}}
+ E K \, \ket{\Psi_{\lBB \rbb} Z'_{\laa \raa}} \\ &
+ F L \, \levi_{\raa \rbb} \levi^{\rCC \rDD} \treelevel{\ket{\Bsi_{\laa \rCC} Y'_{\lBB \rDD}}}
+ F K \, \levi_{\raa \rbb} \levi^{\rCC \rDD} \ket{Y_{\lBB \rCC} \Bsi'_{\laa \rDD}} \\[3mm]
\Smatrix \ket{Z_{\laa \raa} \Bsi'_{\lbb \rBB}} = \ &
+ D L \, \treelevel{\ket{Z_{\laa \raa} \Bsi'_{\lbb \rBB}}}
+ E L \, \treelevel{\ket{Z_{\lbb \raa} \Bsi'_{\laa \rBB}}}
+ D K \, \treelevel{\ket{\Bsi_{\laa \rBB} Z'_{\lbb \raa}}}
+ E K \, \ket{\Bsi_{\lbb \rBB} Z'_{\laa \raa}} \\ &
+ F L \, \levi_{\laa \lbb} \levi^{\lCC \lDD} \treelevel{\ket{\Psi_{\lCC \raa} Y'_{\lDD \rBB}}}
+ F K \, \levi_{\laa \lbb} \levi^{\lCC \lDD} \ket{Y_{\lCC \rBB} \Psi'_{\lDD \raa}} \\[3mm]
\Smatrix \ket{\Psi_{\lAA \raa} Z'_{\lbb \rbb}} = \ &
+ D G \, \treelevel{\ket{\Psi_{\lAA \raa} Z'_{\lbb \rbb}}}
+ D H \, \treelevel{\ket{Z_{\lbb \raa} \Psi'_{\lAA \rbb}}}
+ E G \, \treelevel{\ket{\Psi_{\lAA \rbb} Z'_{\lbb \raa}}}
+ E H \, \ket{Z_{\lbb \rbb} \Psi'_{\lAA \raa}} \\ &
+ F G \, \levi_{\raa \rbb} \levi^{\rCC \rDD} \treelevel{\ket{Y_{\lAA \rCC} \Bsi'_{\lbb \rDD}}}
+ F H \, \levi_{\raa \rbb} \levi^{\rCC \rDD} \ket{\Bsi_{\lbb \rCC} Y'_{\lAA \rDD}} \\[3mm]
\Smatrix \ket{\Bsi_{\laa \rAA} Z'_{\lbb \rbb}} = \ &
+ D G \, \treelevel{\ket{\Bsi_{\laa \rAA} Z'_{\lbb \rbb}}}
+ E G \, \treelevel{\ket{\Bsi_{\lbb \rAA} Z'_{\laa \rbb}}}
+ D H \, \treelevel{\ket{Z_{\laa \rbb} \Bsi'_{\lbb \rAA}}}
+ E H \, \ket{Z_{\lbb \rbb} \Bsi'_{\laa \rAA}} \\ &
+ F G \, \levi_{\laa \lbb} \levi^{\lCC \lDD} \treelevel{\ket{Y_{\lCC \rAA} \Psi'_{\lDD \rbb}}}
+ F H \, \levi_{\laa \lbb} \levi^{\lCC \lDD} \ket{\Psi_{\lCC \rbb} Y'_{\lDD \rAA}}
\end{split}
\]

\bibliographystyle{nb}
\bibliography{twoloop}

\end{document}